\documentclass[journal,a4paper]{IEEEtran}
\usepackage{amsmath,amsfonts}
\usepackage{algorithmic}
\usepackage{algorithm}
\usepackage{array}
\usepackage{textcomp}
\usepackage{stfloats}
\usepackage{url}
\usepackage{cite}
\usepackage{verbatim}
\usepackage[pdftex]{graphicx}
\usepackage{color}
\usepackage{transparent}
\usepackage{import}
\usepackage{subfigure}
\usepackage{balance}
\usepackage{multirow}
\usepackage{cleveref}
\DeclareGraphicsExtensions{.eps}
\usepackage{epstopdf}
\usepackage{epsfig}
\usepackage{gensymb}
\usepackage{mathrsfs,amsmath}
\usepackage{float}
\def\BibTeX{{\rm B\kern-.05em{\sc i\kern-.025em b}\kern-.08em
    T\kern-.1667em\lower.7ex\hbox{E}\kern-.125emX}}
\usepackage{balance}
\usepackage{amsmath,amsfonts,amssymb}
\begin{document}

\title{Permittivity Characterization of Human Skin Based on a Quasi-optical System at Sub-THz}

\author{Bing Xue,~\IEEEmembership{Graduate Student Member,~IEEE}, Juha Tuomela, Katsuyuki Haneda, ~\IEEEmembership{Member,~IEEE}, Clemens Icheln, Juha Ala-Laurinaho
\thanks{paper received April XX, 2023; revised August XX, 2023.}
\thanks{B. Xue, J. Tuomela, K. Haneda, J. Ala-Laurinaho, and C. Icheln are with the Department of Electronics and Nanoengineering, Aalto University-School of Electrical Engineering, Espoo FI-00076, Finland (e-mail: bing.xue@aalto.fi).

The results presented in this paper have been supported by the Academy of Finland -- NSF joint call pilot ``Artificial intelligence and wireless communication technologies", decision \# 345178.}}
\markboth{Journal of \LaTeX\ Class Files,~Vol.~14, No.~8, April~2023}%
{Shell \MakeLowercase{\textit{et al.}}: A Sample Article Using IEEEtran.cls for IEEE Journals}

\maketitle

\begin{abstract}
This paper introduces a novel approach to experimentally characterize effective human skin permittivity at sub-Terahertz (sub-THz) frequencies, specifically from $140$~to $210$~GHz, utilizing a quasi-optical measurement system. To ensure accurate measurement of the reflection coefficients of human skin, a planar, rigid, and thick reference plate with a low-loss dielectric is utilized to flatten the human skin surface. A permittivity characterization method is proposed to reduce permittivity
estimation deviations resulting from the pressure effects on the phase displacements of skins under the measurements but also to ensure repeatability of the measurement. In practical permittivity characterizations, the complex permittivities of the finger, palm, and arm of seven volunteers show small standard deviations for the repeated measurements, respectively, while those show significant variations across different regions of the skins and for different persons. The proposed measurement system holds significant potential for future skin permittivity estimation in sub-THz bands, facilitating further studies on human-electromagnetic-wave interactions based on the measured permittivity values.
\end{abstract}


\section{Introduction}
\IEEEPARstart{K}{nowledge} of electrical properties of human tissues is of importance in various fields, including the medicine~\cite{pickwell2004vivo,feldman2009electromagnetic,zakharov2009full,wilmink2011Development,sasaki2017monte}, biology~\cite{betzalel2017modeling,Wang2019Wideband,Baksheeva2021The,benova2021human}, dosimetry~\cite{rashed2019human,benova2021human,Standard2019}, and wireless communications and sensing~\cite{Zhekov2019,gao2018towards,zhekov2020,christ2021reflection,nor2022effect}. Among varying tissue types, knowledge of human skin permittivity enables understanding of human body status~\cite{zakharov2009full}, designing biosensors for human imaging~\cite{Wang2019Wideband,Baksheeva2021The}, and studying power-efficient radio communication devices and physical layer schemes against absorption and blockage of radio signals by a human body~\cite{Lauri2021,Xue2022,nor2022effect}. Recent research at sub-terahertz (sub-THz), e.g., radio communications~\cite{Erturk2018A,nor2022effect,Zhang2023Measure, Kwon2023Development} and human body imaging~\cite{Wu2023MIMO, Wu2022Terahertz}, consider the radio frequency (RF) band between 100 GHz and 300 GHz, where the knowledge of electrical properties of human tissues in that RF band is required. 
Despite papers reporting human skin permittivity characteristics at sub-THz, e.g.,~\cite{pickwell2004vivo,feldman2009electromagnetic,wilmink2011Development,xue2024Human} and references therein, the insights provided by the papers do not suffice for studies of electromagnetic field effects on human body at the RF band. Models of effective permittivity of different human body parts such as fingers, palms, arms, and torso, and their variation across people, are limited. These body parts are particularly relevant to the study because they affect e.g., the radiation characteristics of mobile terminals when users are actively engaged with their devices. 

At sub-THz, existing permittivity characterization methods~\cite{Kazemipour2015Design, Yashchyshyn2018A, Zhu2021complex, aliouane2022material,xue2024thick} mainly focus on planar low-loss dielectric materials such as substrates for printed circuit board, crystal and glass. They are not suitable for skin permittivity measurements since the soft and sweat-sensitive nature of human skin permittivity presents challenges in obtaining accurate permittivity estimates. 
The methods for human skin permittivity estimation use coaxial cables and waveguide openings as a probe~\cite{Zhekov2019, gao2018towards, christ2021reflection, xue2024Human}, on which a subject human places a body part under test. Thin flat materials are introduced on the probe-body interface or low-loss material infills the waveguide openings so that the human skin becomes a flat medium touching the interface and does not protrude into the probe.  
The best permittivity-estimate uncertainty of $<\pm 1~\%$ was reported in~\cite{xue2024Human} for finger-skin permittivity over $140$-$220$~GHz RF. However, this method can only measure the permittivity of tissues touching the very small area of WR5 waveguide opening, i.e., $<0.9~\rm mm^2$. More relevant permittivity estimates for a study of electromagnetic effects on human body in communications, dosimetry and body sensing would be those of different body parts, as extensively studied in the below-6 GHz and 5G millimeter-wave frequencies~\cite{Sasaki2015,gao2018towards,Zhekov2019,christ2021reflection,xue2024Human}. 

The novel effective permittivity measurement method in this paper is based on a quasi-optical system, which is widely used for RF between 100~GHz and 300~GHz~\cite{Kazemipour2015Design, Yashchyshyn2018A, Zhu2021complex, aliouane2022material,xue2024thick}. It excites a quasi-plane wave with a wide beam, which can cover relatively large body parts. Taking advantage of the wisdom of probe-based measurement methods~\cite{Zhekov2019, gao2018towards, christ2021reflection, xue2024Human}, the proposed method flattens the skin surface using a reference low-loss rigid plate. 
Different from the probe-based methods, permittivity estimation in free space requires precise placement and alignment of phase reference planes across measurements~\cite{Kazemipour2015Design,Yashchyshyn2018A}. However, consistent displacement and alignment across measurements become a particular challenge at sub-THz due to its short wavelength. While precision displacement platforms help the purpose, they 
still introduce minor phase deviations. When human subjects having non-repeatable behavior fix their body parts on the reference plate by applying strength, variation of the strength leads to phase deviations and degrades measurement repeatability. In light of this challenge, our novel permittivity estimation method corrects the displacement and alignment by phase retrieval of an incident plane wave at the interfaces between the reference plate and human tissue or air, allowing improved repeatability of measurements.
We show in this paper that the uncertainty of skin permittivity estimates for a large body part is comparable to that of the best probe-based method in~\cite{xue2024Human}.

The remaining sections of the paper are organized as follows: Section~\ref{sec:HumanSkin} reviews the layered model of human skin to justify the need for partition modeling of the human body's permittivity. In Section~\ref{sec:systemdesign}, our approach of permittivity estimation is derived, where the phase correction method for displacement and misalignment of the reference plate is elaborated. Full-wave simulations are implemented to validate the proposed algorithms. Section~\ref{sec:SystemSetup} presents the quasi-optical system for human permittivity characterizations and its calibration. The selection criteria for the reference plate are discussed.
In Section~\ref{sec:MeasurementResults}, the effectiveness of the proposed measurement system is experimentally demonstrated. The permittivities of the finger, palm, and arm of seven volunteers are presented. Finally, Section~\ref{sec:conclusion} provides key findings of the present study and their implications.

\section{Variation of Skin Permittivity across Body Parts}
\label{sec:HumanSkin}
\begin{figure}[htbp]
    \centering
	\includegraphics[width=1.0\linewidth]{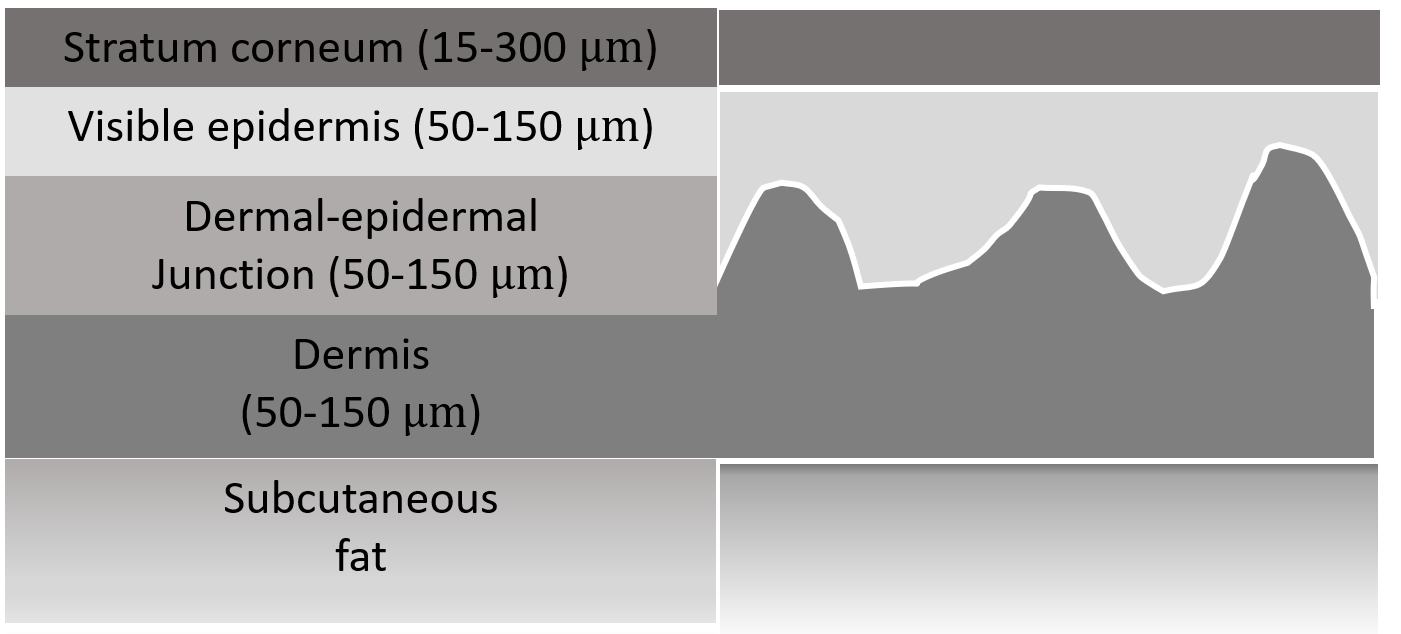}
	\caption{Multi-layer model of human skin. The thickness of subcutaneous fat is considered large compared to the skin depth.}
	\label{fig:skinlayer}
\end{figure}

Above 30 GHz, the electromagnetic wave's skin depth in human skin is within $2$~mm. Consequently, the interaction between humans and electromagnetic waves can be described as a skin-electromagnetic wave interaction~\cite{Xue2022}. The structure of human skin can be represented by multiple layers, as depicted in Fig.~\ref{fig:skinlayer}. When operating below 60~GHz, it is reasonable to use the human skin permittivity value specified in~\cite{Gabriel1996} as an averaged value over persons. This is because the variation in permittivity among individuals or different parts of a person has minimal impact on electromagnetic radiation below 60~GHz~\cite{Zhekov2019}.

However, above 100 GHz, human skin permittivities may differ from that below 60~GHz mainly due to the influence of skin moisture and the thickness of each skin layer. The thickness of the stratum corneum layer varies between $15~\rm \mu m$ on the arm and $300~\rm \mu m$ on the fingers~\cite{zakharov2009full}. The thickness of the stratum corneum is comparable to the wavelength of sub-THz frequencies. This leads to non-negligible variations of skin permittivity over the human body. Additionally, human skin is soft and sometimes sweaty, so skin permittivity characteristics vary between persons~\cite{pickwell2004vivo}. Consequently, at sub-THz, it is necessary to evaluate variations of skin permittivities across different parts of human bodies. 

\section{Permittivity characterization Method}
\label{sec:systemdesign}

\subsection{Theoretical Foundation}
\label{sec:Theoretical}
\begin{figure}[htbp]
    \centering
 \subfigure[]{
	\includegraphics[width=0.75\linewidth]{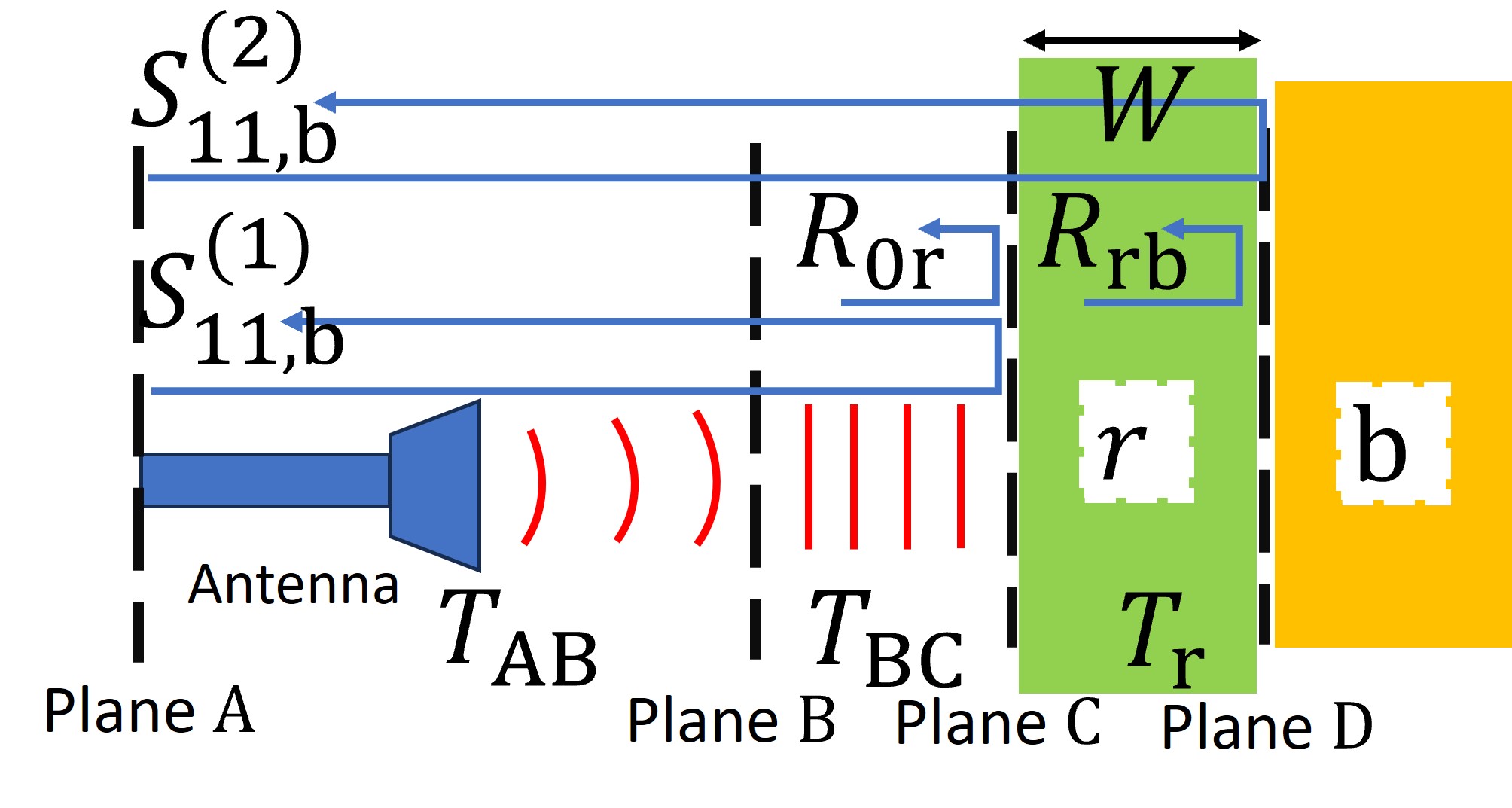}\label{fig:schematic2}}
 	\subfigure[]{
	\includegraphics[width=0.75\linewidth]{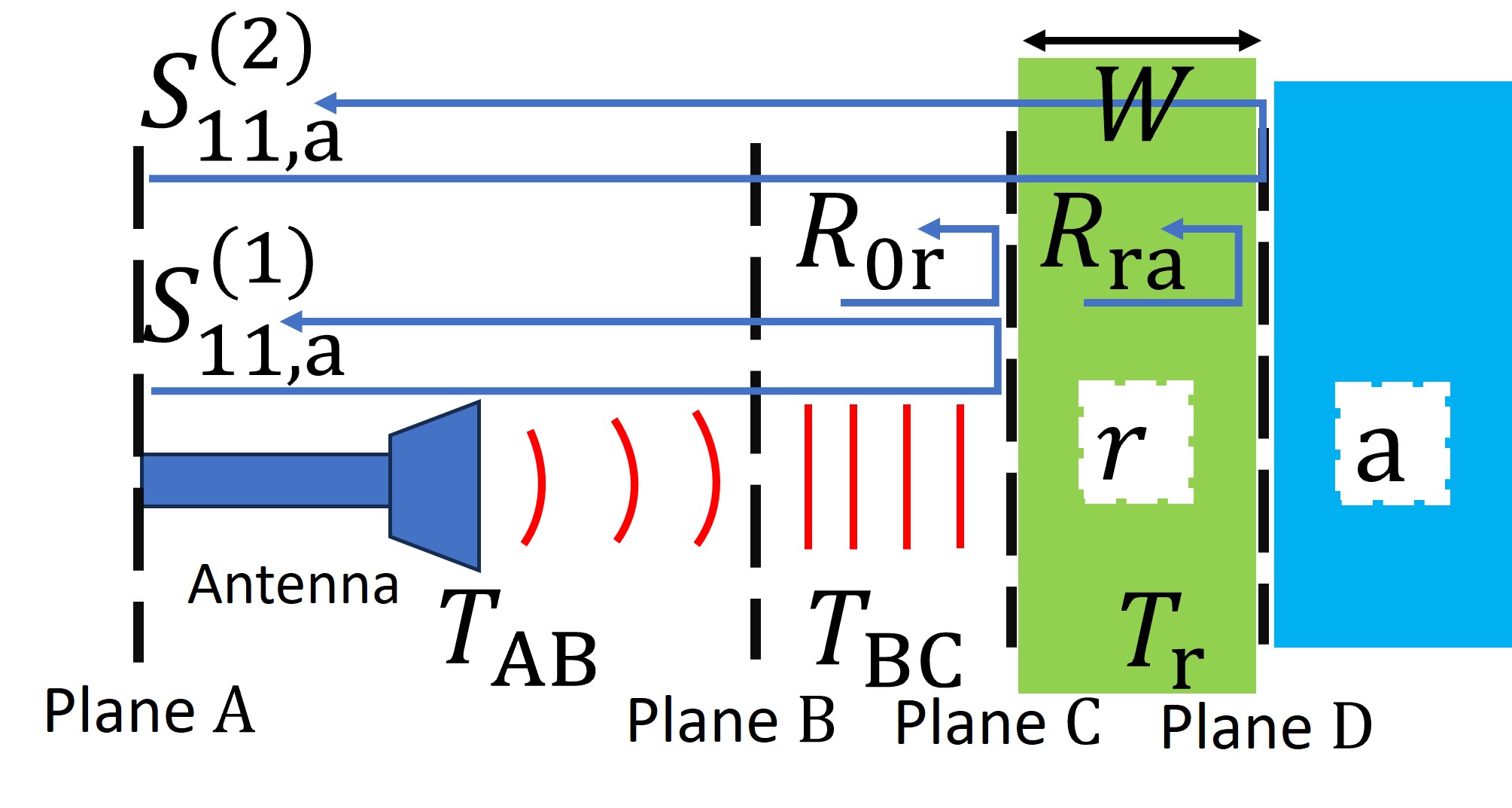}\label{fig:schematic2a}}
	\caption{Schematic of permittivity measurement for (a) material under test [$\mathbf  b$] and (b) known material [$\mathbf  a$]. The green rectangle represents the reference plate $\mathbf r$; the yellow rectangle is the material under test [$\mathbf  b$]; the blue rectangle is the unknown material [$\mathbf  a$].}
	\label{fig:permmeasurement}
\end{figure}

The measurement method is depicted in Fig.~\ref{fig:permmeasurement}. The reflection coefficient of an illuminating plane wave $S^{(2)}_{11}$ on the further interface (Plane D) of the reference plate, indicated by the green color, can be expressed as follows:
\begin{equation}
\begin{aligned}
S^{(2)}_{11} &= T_{\rm AB}T_{\rm BC}\left(1+R_{\rm 0r}\right)T_{\rm r}R_{\rm rb}T_{\rm r}\left(1+R_{\rm r0}\right)T_{\rm CB}T_{\rm BA}\,
\\&= T_{\rm AB}^{2}T_{\rm BC}^{2}\left(1-R_{\rm 0r}^{2}\right)T_{\rm r}^{2}R_{\rm rb},
\label{eq:S211}
\end{aligned}
\end{equation}
where $T_{\rm AB}=T_{\rm BA}$ represents the transmission coefficient between the extender of a vector network analyzer (VNA) calibration plane (Plane A) and the reference plane in the plane wave zone (Plane B), and $T_{\rm BC}=T_{\rm CB}$ represents the transmission coefficient between Plane B and the first surface of the reference plate (Plane C). 
The transmission coefficient $T_{\rm r}$ of a plane wave through the reference plate $\mathbf r$ is given by:
\begin{equation}
T_{\rm r} = \exp{\left(-{\rm i}k_0 W\sqrt{\varepsilon_{\rm r}}\right)}
\label{eq:Trans}
\end{equation}
where $\varepsilon_{\rm r}$ represents the complex relative permittivity of the reference plate $\mathbf r$, which we assume to be {\it known} here;
$W$ is its thickness, and $k_0$ is the wave number in free space. Finally, the reflection coefficient $R_{\rm rb}$ arises when a normally incident wave on the reference plate [$\mathbf r$] is reflected by the material [$\mathbf b$] in Fig.~\ref{fig:schematic2} and can be calculated using the expression:
\begin{equation}
R_{\rm rb} = \frac{\sqrt{\varepsilon_{\rm r}} -\sqrt{\varepsilon_{\rm b}}}{\sqrt{\varepsilon_{\rm r}}+\sqrt{\varepsilon_{\rm b}}}.
\label{eq:Reflect}
\end{equation}
Similarly, $R_{\rm 0r} = -R_{\rm r0}$ represents the reflection coefficient when an incident wave is reflected from the reference plate [$\mathbf  r$] to free space. 

To estimate the permittivity of the material under test (MUT) $\varepsilon_{\rm b}$ using~\eqref{eq:S211}, we need to know the values of $T_{\rm AB}$, $T_{\rm BC}$, and $W$. The latter set can be estimated by replacing the MUT [$\mathbf  b$] with a material of a known permittivity ($\varepsilon_{\rm a}$) with a flat surface in Fig.~\ref{fig:schematic2}, and measuring reflection coefficients
\begin{equation}
\begin{aligned}
S^{(2)}_{11,a} = T_{\rm AB}^{2}T_{\rm BC}^{2}\left(1-R_{\rm 0r}^{2}\right)T_{\rm r}^{2}R_{\rm ra}
\label{eq:S11am}
\end{aligned}
\end{equation}
for the material with the known permittivity and
\begin{equation}
\begin{aligned}
S^{(2)}_{11,b} = T_{\rm AB}^{2}T_{\rm BC}^{2}\left(1-R_{\rm 0r}^{2}\right)T_{\rm r}^{2}R_{\rm rb}
\label{eq:S11bm}
\end{aligned}
\end{equation}
for the MUT. It should be noted that the measurement environment and the placement and alignment of the reference plate [$\mathbf  r$] must precisely remain unchanged during the two measurements so that the term $T_{\rm AB}^{2}T_{\rm BC}^{2}\left(1-R_{\rm 0r}^{2}\right)T_{\rm r}^{2}$ is identical across~\eqref{eq:S11am} and~\eqref{eq:S11bm}~\cite{Yashchyshyn2018A}. 

Solving~\eqref{eq:Reflect},~\eqref{eq:S11am}, and~\eqref{eq:S11bm} yields an estimate of $\varepsilon_{\rm b}$ as:
\begin{equation}
\varepsilon_{\rm b} = \varepsilon_{\rm r} \left(\frac{S^{(2)}_{\rm 11,a} - R_{\rm ra}S^{(2)}_{\rm 11,b}}{S^{(2)}_{\rm 11,a} + R_{\rm ra}S^{(2)}_{\rm 11,b}}\right)^2.
\label{eq:meas}
\end{equation}
In practice, air ($\varepsilon_{\rm a} = 1$) is often chosen for convenience. Therefore, analogous to~\eqref{eq:Reflect}, $R_{\rm ra}$ can be represented as
\begin{equation}
R_{\rm ra} = \frac{\sqrt{\varepsilon_{\rm r}} -1}{\sqrt{\varepsilon_{\rm r}}+1}.
\label{eq:airreflec}
\end{equation}

It is noted that the thickness information for the reference plate is not needed in~\eqref{eq:meas}, which can reduce the uncertainty of the permittivity estimate due to the deviation in the thickness estimation of the reference plate. 
\subsection{Phase Correction Method}
\label{sec:Compensation}
The identity of the term $T_{\rm AB}^{2}T_{\rm BC}^{2}\left(1-R_{\rm 0r}^{2}\right)T_{\rm r}^{2}$ across the two measurements~\eqref{eq:S11am} and~\eqref{eq:S11bm} is influenced by changes of the measurement environment, e.g, the way a human subject attaches its body parts, corresponding to MUT, on the reference plate. In order to ensure the utmost identity of the term, we introduce a phase correction method.
\begin{figure}[htbp]
    \centering
	\subfigure[]{
	\includegraphics[width=0.85\linewidth]{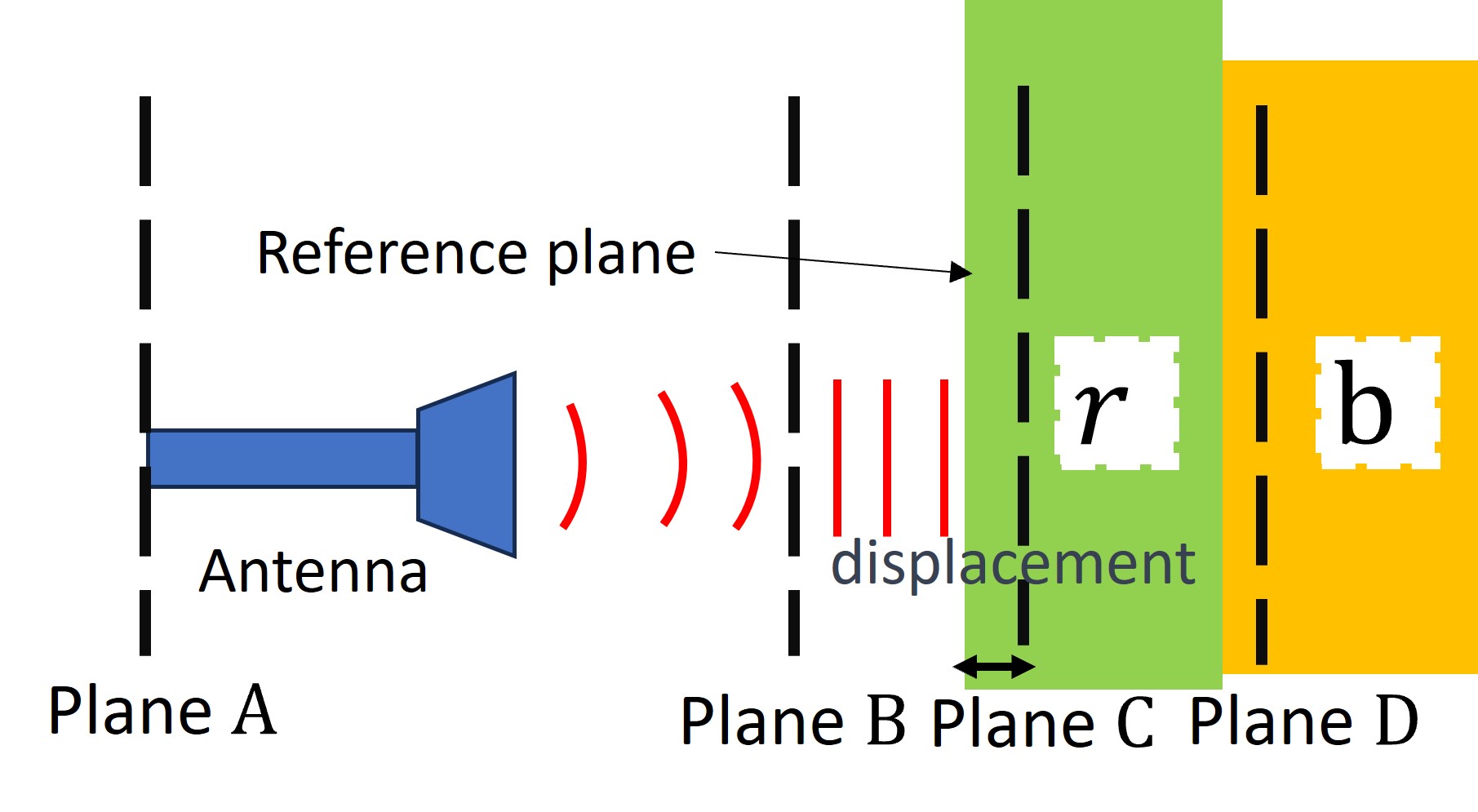}\label{fig:positionmismatch}}
 	\subfigure[]{
	\includegraphics[width=0.85\linewidth]{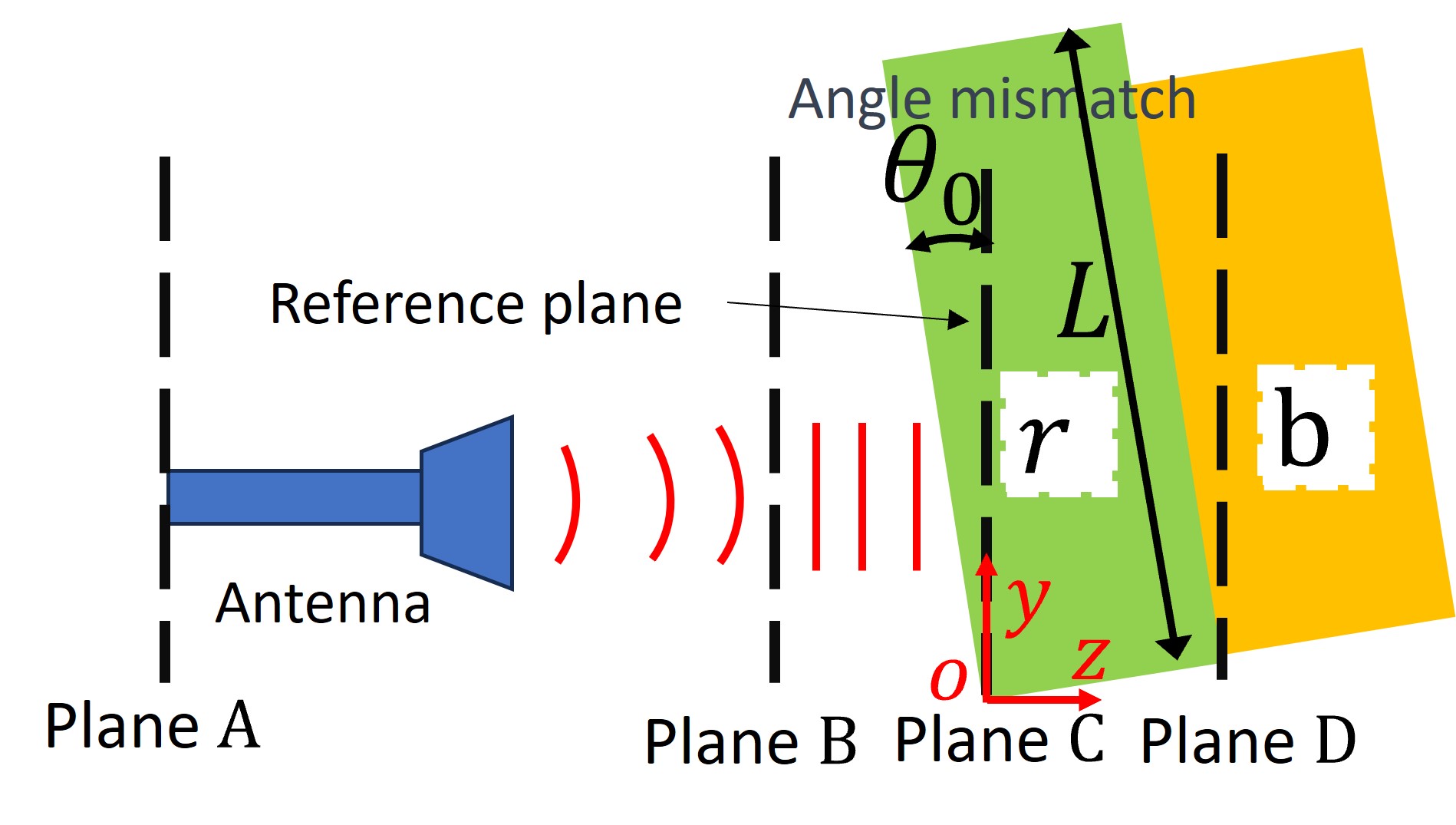}\label{fig:anglemismatch}}
  	\subfigure[]{
	\includegraphics[width=0.5\linewidth]{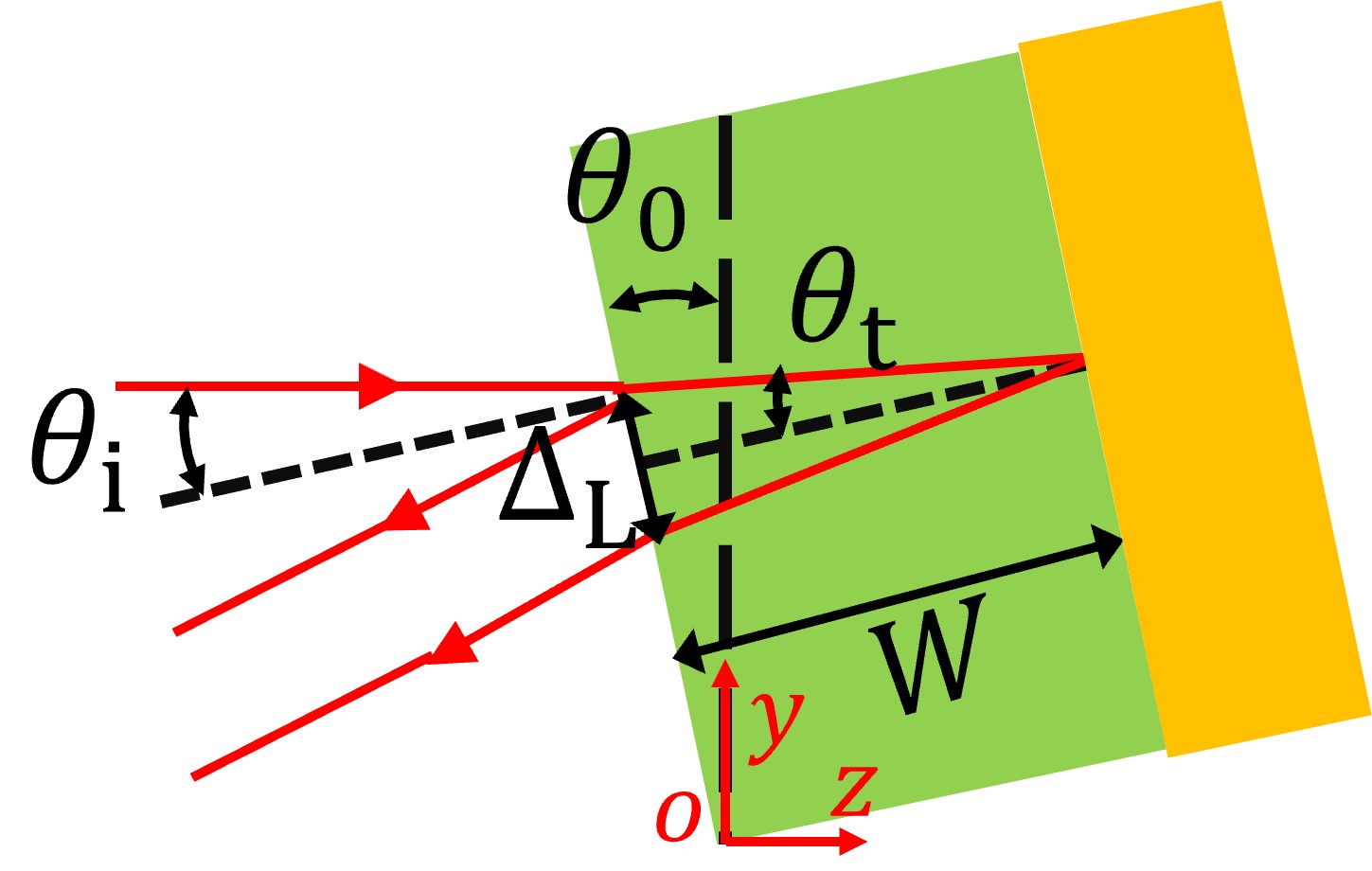}\label{fig:angles}}
	\caption{Diagrams of a) displacement and b) inclination of the reference plate. c) A local view of the reference plate for b).}
	\label{fig:mismatch}
\end{figure}
\subsubsection{Displacement of the Reference Plate}
One issue is the displacement of the reference plate between air and MUT measurements, as illustrated in Fig.~\ref{fig:positionmismatch}. Let us denote $A = T_{\rm AB}T_{\rm BC}$, then~\eqref{eq:S11am} and~\eqref{eq:S11bm} becomes
\begin{eqnarray}
\label{eq:S211r1}
S^{(2)}_{\rm 11,a} &= & A^2_{\rm a}\left(1-R_{\rm 0r}^{2}\right)T_{\rm r}^{2}R_{\rm ra}, \\
S^{(2)}_{\rm 11,b} & = & A^2_{\rm b}\left(1-R_{\rm 0r}^{2}\right)T_{\rm r}^{2}R_{\rm rb},
\label{eq:S211b1}
\end{eqnarray}
where $A_{\rm a}$ represents the transmission coefficient of incident and reflected waves over free-space including Ohmic losses of the antenna when the known material [$\mathbf  a$] is attached to the reference plate [$\mathbf  r$]. While $A_{\rm b}$ is the same transmission coefficient when the MUT [$\mathbf  b$] is on the same reference plate [$\mathbf  r$] but subject to displacement compared to the first measurement, leading to a possible difference of $A_{\rm b}$ from $A_{\rm a}$. The difference can be estimated by observing plane wave reflections on the near surface of the reference plate (Plane C). The measured reflection coefficients can be represented as
\begin{eqnarray}
S^{(1)}_{\rm 11,a}&= & A^2_{\rm a}R_{\rm 0r},\label{eq:S111r}\\ 
S^{(1)}_{\rm 11,b}&= &A^2_{\rm b}R_{\rm 0r}.\label{eq:S111b}
\end{eqnarray}
Solving~\eqref{eq:S211r1},~\eqref{eq:S211b1},~\eqref{eq:S111r} and~\eqref{eq:S111b} gives the elaborated relative permittivity estimate of MUT under presence of the displacement as
\begin{equation}
\varepsilon_{\rm b} = \varepsilon_{\rm r} \left(\frac{S^{(1)}_{\rm 11,b}S^{(2)}_{\rm 11,a} - R_{\rm ra}S^{(1)}_{\rm 11,a}S^{(2)}_{\rm 11,b}}{S^{(1)}_{\rm 11,b}S^{(2)}_{\rm 11,a} + R_{\rm ra}S^{(1)}_{\rm 11,a}S^{(2)}_{\rm 11,b}}\right)^2.
\label{eq:meas1}
\end{equation}

\subsubsection{Inclination of the Reference Plate}
The human subjects may also cause the inclination of the reference plate, as shown in Fig.~\ref{fig:anglemismatch}. The reference plate is assumed to be perpendicular to the ground during measurement with the known material parameter $\varepsilon_{\rm a}$. However, when the MUT such as a human subject touches the reference plate, it may be inclined by angle $\theta_{0}$ to the ground. Consequently, the measured reflection coefficients of a plane wave on the near and far surfaces of the reference plate, denoted as $S^{(1)}_{\rm 11,b}$ and $S^{(2)}_{\rm 11,b}$, are written as
\begin{eqnarray}
\label{eq:S111b2}
S^{(1)}_{\rm 11,b} & = & {A_{\rm b}}R_{\rm 0r,b}{A_{\rm b}^{\prime}},\\
\label{eq:S211b2}
S^{(2)}_{\rm 11,b} & = & {A_{\rm b}}\left(1-R_{\rm 0r,b}^{2}\right)T_{\rm r,b}^{2}R_{\rm rb} {A_{\rm b}^{\prime\prime}},
\end{eqnarray}
where $T_{\rm r,b}$ is the transmission coefficient in the reference plate when measured with the MUT [$\mathbf b$] with inclination happening. $R_{\rm 0r,b}$ is a reflection coefficient of a plane wave on a surface of the inclined reference plate. For the horizontally polarized electric field of the incident wave with respect to the ground, it can be expressed as
\begin{equation}
R_{\rm 0r,b} = \frac{\cos{\theta_{\rm i}} - \sqrt{\varepsilon_{\rm r} - \sin^{2}{\theta_{\rm i}}} }{\cos{\theta_{\rm i}} + \sqrt{\varepsilon_{\rm r} - \sin^{2}{\theta_{\rm i}}}},
\label{eq:inclineb}
\end{equation}
while the reflection coefficient of the plane wave on the interface between the reference plate and MUT, $R_{\rm rb}$, now is different from~\eqref{eq:Reflect} due to the inclination of the reference plate as
\begin{equation}
R_{\rm rb} = \frac{\cos{\theta_{\rm t}} - \sqrt{\frac{\varepsilon_{\rm b}}{\varepsilon_{\rm r}} - \sin^{2}{\theta_{\rm t}}} }{\cos{\theta_{\rm t}} + \sqrt{\frac{\varepsilon_{\rm b}}{\varepsilon_{\rm r}} - \sin^{2}{\theta_{\rm t}}}},
\label{eq:newrb}
\end{equation}
where $\theta_{\rm t}$ represents the refraction angle of the plane wave satisfying the Snell's law $\sin{\theta_{\rm i}} = \sqrt{\varepsilon_{\rm r}}\sin{\theta_{\rm t}}$; \eqref{eq:inclineb} and~\eqref{eq:newrb} for the vertically polarized electric field of the incident wave can be found in~\cite{pozar2011}. 

In~\eqref{eq:S111b2}, ${A_{\rm b}}$ is an incident wave's free-space transmission coefficient for unknown material [$\mathbf  b$]; ${A_{\rm b}^{\prime}}$ represents the same but subject to the inclination of the reference plate due to the presence of the for the MUT [$\mathbf  b$]. In~\eqref{eq:S211b2}, ${A_{\rm b}^{\prime\prime}}$ is the transmission coefficients of the plane wave reflected from the far surface under the same condition to ${A_{\rm b}^{\prime}}$. Because of the inclination of the reference plate, $A_{\rm b} \neq A_{\rm b}^{\prime} \neq A_{\rm b}^{\prime\prime}$. The geometry of inclination is depicted in Fig.~\ref{fig:angles}, where $\theta_{\rm i}$ is the incident angle from the air to the reference plate with respect to the reference plate's surface; $\theta_{\rm i} = \theta_{0}$ when the inclined angle of the reference plate is $\theta_0$; 
$A_{\rm b}^{\prime} \neq A_{\rm b}^{\prime\prime}$ mainly because of a difference of free-space propagation distance for the two reflected waves by 
$\Delta_{\rm L} \sin \theta_0= 2W\sin\theta_0\tan{\theta_{\rm t}}$ where $\Delta_{\rm L}$ is defined in Fig.~\ref{fig:angles}. Vector $\mathbf{k_0^\prime}=[0, k_0\sin(2\theta_{\rm 0}), k_0\cos(2\theta_{\rm 0})]$ is the free-space wave number of the two reflected waves after inclination. Consequently, we can estimate ${A_{\rm b}^{\prime}}/{A_{\rm b}^{\prime\prime}} \approx \exp{\left(-{\rm i} k_{0}\cos(2\theta_{\rm 0}) \Delta_{\rm L} \sin \theta_0\right)} = \exp{\left(-2{\rm i} k_{0} W\sin\theta_{\rm 0}\tan\theta_{\rm t}\cos(2\theta_{\rm 0})\right)}$.

Using~\eqref{eq:S211r1},~\eqref{eq:S111r},~\eqref{eq:S111b2} and~\eqref{eq:S211b2}, $R_{\rm rb}$ can be represented as
\begin{equation}
\begin{aligned}
R_{\rm rb}  &= R_{\rm ra}\frac{S^{(2)}_{\rm 11,b}}{S^{(2)}_{\rm 11,a}}\frac{S^{(1)}_{\rm 11,a}}{S^{(1)}_{\rm 11,b}} \frac{1-R_{\rm 0r,a}^{2}}{1-R_{\rm 0r,b}^{2}} \frac{R_{\rm 0r,b}}{R_{\rm 0r,a}} \frac{T_{\rm r,a}^{2}}{T_{\rm r,b}^{2}}\\ &\cdot\exp{\left(-2{\rm i} k_{0} W\sin\theta_{\rm 0}\tan\theta_{\rm t}\cos(2\theta_{\rm 0})\right)}.
\label{eq:measq}
\end{aligned}
\end{equation}
where $T_{\rm r,a}$ is the transmission coefficient in the reference plate when measured with known material [$\mathbf a$]. We can derive:
\begin{equation}
\frac{T_{\rm r,a}^{2}}{T_{\rm r,b}^{2}} = \exp{\left(2{\rm i} k_{0} W\sqrt{\varepsilon_{\rm r}}\left(\frac{1}{\cos{\theta_{\rm t}}}-1\right)\right)}
\label{eq:meas3}
\end{equation}
Assume that the known material is air ($\varepsilon_{\rm a}=1$) for convenience and that there is no inclination when measuring the known material [$\mathbf a$], that is,
$R_{\rm 0r,a} = (1-\sqrt{\varepsilon_{\rm r}} )/(1+\sqrt{\varepsilon_{\rm r}})$.

To estimate the inclination angle $\theta_{0}$, it can be assumed that the displacement of the reference plate is negligible. Thus, the phase change of a reflected plane wave on the reference plate due to the inclination can be estimated as
\begin{equation}
\Delta_{\rm A} \approx \angle{\left(\frac{S^{(1)}_{\rm 11,a}}{S^{(1)}_{\rm 11,b}}\right) + 2n\pi},
\label{eq:extract2}
\end{equation}
where $\angle{\left(\cdot\right)}$ is the operator to obtain the phase of a complex value, and $n$ is a natural number;  $n = 0$ for typical inclination angles we consider. 

\subsubsection{Correction of Displacement and Inclination of the Reference Plate}
If the reference plate is subject to both inclination and displacement and we assume that the known material is air, we can also measure $S^{(1)}_{\rm 11,a}$ before and after measuring $S^{(1)}_{\rm 11,b}$, denoted as $S^{(1,1)}_{\rm 11,a}$ and $S^{(1,2)}_{\rm 11,a}$ respectively. The phase shift due to the reference plate displacement $\Delta_{\rm P}$ can be, therefore, modelled by 
\begin{equation}
\Delta_{P} \approx \angle{\left(\frac{S^{(1,1)}_{\rm 11,a}}{S^{(1,2)}_{\rm 11,a}}\right)}.
\label{eq:extract2_3}
\end{equation}
Accordingly, we can also obtain $\Delta_{\rm A}$:
\begin{equation}
\begin{aligned}
\Delta_{\rm A} &\approx \angle{\left(\frac{S^{(1,2)}_{\rm 11,a}}{S^{(1)}_{\rm 11,b}}\right)} \\&\approx 2k_0\sin\theta_0\left[(1-\sin^2\theta_0)L_{\rm c} + \sin\theta_0d_0\right]. \label{eq:extract2_2}
\end{aligned}
\end{equation}
where $L_{\rm c}$ is the height of the MUT from the ground to the plane-wave illumination area center, The ground is $xoz$ defined in Fig.~\ref{fig:anglemismatch}. $d_0$ is the free space propagation distance.
Using these estimates, $\theta_{\rm 0}$ can be approximated by solving~\eqref{eq:extract2_2}. $L_{\rm c}$ can be obtained by rulers, and $d_0$ can be extracted by measured $S^{(1)}_{\rm 11,a}$ in time domain. In practice, $(1-\sin^2\theta_0)L_{\rm c}\gg\sin\theta_0d_0$ can be achieved. In this way, $\theta_{\rm i}$ can be roughly estimated by
\begin{equation}
\theta_{\rm i}=\theta_{\rm 0} \approx\arcsin{\left(\frac{\Delta_{\rm A}}{2k_{0}L_{\rm c}}\right)}.
\label{eq:extract3}
\end{equation}
Solving~\eqref{eq:newrb} and~\eqref{eq:measq}, the elaborated relative permittivity estimate considering the correction of both displacement and inclination effects of the reference plane becomes
\begin{equation}
\varepsilon_{\rm b} = \varepsilon_{\rm r} \left(\frac{1-R_{\rm rb}}{1+R_{\rm rb}}\right)^2 +\frac{\Delta_{\rm A}}{k_{0}L_{\rm c}}\frac{4R_{\rm rb}}{\left(1+R_{\rm rb}\right)^2}.
\label{eq:meas2}
\end{equation}
where $R_{\rm rb}$ is calculated by~\eqref{eq:measq}.

\subsection{Multiplicative Noise and Drift Correction}
Multiplicative noises or measurement deviations due to temperature drift of the nonlinear components in the VNA can be modeled as
\begin{equation}
{S^\prime}_{11} = KS_{11}.
\label{eq:uncertaintyS}
\end{equation}
where $K\in\mathbb{C}$ represents the modulation coefficient of the multiplicative noises. Consequently, the measured reflection coefficients of the first and second bounces from the reference plate [$\mathbf  r$] can be written as ${S^\prime}_{11}^{(1)} = KS_{11}^{(1)}$ and ${S^\prime}_{11}^{(2)} = KS_{11}^{(2)}$. Therefore, it is evident that the multiplicative noise and drift can also be compensated to some extent using~\eqref{eq:meas1} or~\eqref{eq:meas2}.

\subsection{Extraction of Reflection Coefficients}
\label{sec:Extraction}
The measured S-parameter can be modeled as
\begin{equation}
\begin{aligned}
\hat{S}_{\rm 11}(f)  &= {S}_{\rm 11}^{(1)}(f) + {S}_{\rm 11}^{(2)}(f) + \sum_{i=3}^\infty{{S}_{\rm 11}^{(i)}(f)} \\&+ \Gamma_{\rm ant}(f) + \Gamma_{\rm env}(f) + N_0(f),
\label{eq:Sparameter}
\end{aligned}
\end{equation}
where ${S}_{\rm 11}^{(i)}(f)$ ($i>2$) represents the third and beyond bounces from the reference plate and the MUT; $\Gamma_{\rm ant}(f)$ corresponds to sum of feeding mismatching and antenna-free-space mismatching; $\Gamma_{\rm env}(f)$ is the reflection coefficients due to surrounding environment; $N_0(f)$ is the system noise. By using the inverse Fourier transform, the time-domain system response can be written as
\begin{equation}
\begin{aligned}
\hat{H}_{11}(t) &= \mathcal{F}^{-1}\left[\hat{S}_{11}(f)\right]  \\&= {h}_{11}^{(1)}(t-t_1) + {h}_{11}^{(2)}(t-t_2) \\&+ \sum_{i=3}^\infty{{h}_{11}^{(i)}(t-t_i)} \\&+ \gamma_{\rm ant}(t-t_{\rm a}) + \gamma_{\rm env}(t) + n_0(t).
\label{eq:Timeresponse} 
\end{aligned}
\end{equation}

In engineering applications, the inverse discrete Fourier transform (IDFT) is often used to obtain the response in the time domain, where the resolution in the time domain is $\delta_t = {1}/{B_0}$, and $B_0$ is the measurement frequency range. Assume that $D$ is an effective time-bin window of the time-domain response. In this permittivity measurement method, the time difference $t_2 - t_1$ should be much larger than $\delta_t $ to separate ${S}_{11}^{(1)}(f)$ and ${S}_{11}^{(2)}(f)$. Thus, we have the condition:
\begin{equation}
t_2 - t_1 > D
\label{eq:timediff}
\end{equation}

Therefore, the thickness and permittivity of the reference plate should satisfy:
\begin{equation}
\sqrt{\varepsilon_{\rm r}}W > \frac{1}{2}{\rm c_0} D,
\label{eq:thickE} 
\end{equation}
where ${\rm c_0}$ is the speed of light in free space. The Blackman window and the Kaiser window are commonly used to isolate the desired signals from the background noise and interference due to their low secondary lobes in the frequency domain. The width of the window needs to preserve the desired signal while excluding the interference signals. Therefore, the window width ($W_{\rm w}$) should be wide enough but exclude the time response of the interference signals. One satisfying window width is
\begin{equation}
2D< W_{\rm w} < 2(t_2 - t_1- \frac{D}{2}).
\label{eq:black} 
\end{equation}
By this formula, we can derive $t_2-t_1 > 3D/2$, which proposes a condition that the thickness and/or dielectric constant of the reference plate should be large enough to satisfy~\eqref{eq:black}.

For materials with low dispersion characteristics in measurable frequency bandwidth,i.e., $90$~GHz, reflected signals typically fall into a ${D} \approx 12\delta_t$ time-bin window normally centered at the $7_{\rm th}$ time-bin based on the IDFT. However, antennas can further broaden the time-domain response because of their frequency dispersion. In particular, corrugated horn antennas can generate Gaussian-like beams, and they are commonly used in quasi-optical systems at sub-THz frequencies. However, the group delay of each frequency is not constant due to the groove structures, resulting in a longer time-domain response. When such antennas are used, the response length in the time domain becomes ${D_{\rm cor}} \approx 24\delta_t$ normally centered at the $7_{\rm th}$ time bin based on experimental observations.

Moreover, part of $\Gamma_{\rm env}(f)$ may overlap with ${S}_{11}^{(1)}(f)$ and ${S}_{11}^{(2)}(f)$. One approach to reduce the environmental influences on the desired signals is to measure the environmental reflections ($\hat{S}_{11,0}$), which is a good estimate of $\Gamma_{\rm ant}(f) + \Gamma_{\rm env}(f)$. The formula can be written as
\begin{equation}
\begin{aligned}
\bar{S}_{11,u} &= \hat{S}_{11,u} - \hat{S}_{11,0} \\&= {S}_{11}^{(1)}(f) + {S}_{11}^{(2)}(f) + \sum_{i=3}^\infty{{S}_{11}^{(i)}(f)} + N_0(f),
\label{eq:newSpa} 
\end{aligned}
\end{equation}
where $\hat{S}_{11,u}$ ($u={\rm a~or~b}$) represents the measured S-parameters of the MUT [$\mathbf  b$] or the known-permittivity material [$\mathbf  a$]. $\bar{S}_{11,u}$ represented the measured S-parameters without $\Gamma_{\rm ant}(f) + \Gamma_{\rm env}(f)$.

\subsection{Limitation Analysis}
In general, $|{S}_{11}^{(1)}(f)|$ and $|{S}_{11}^{(2)}(f)|$ should be larger than the system noise $\left|N_0(f)\right|$. When the MUT has a permittivity similar to the reference plate [$\mathbf  r$], $\left|{S}_{11,b}^{(2)}(f)\right|$ will be close to the noise level. Therefore, we can establish the following conditions:
\begin{equation}
\begin{aligned}
\left|{S}_{\rm 11,b}^{(2)}(f)\right| = \left|A_{\rm b}\left(1-R_{\rm 0r}^{2}\right)T_{\rm r}^{2}R_{\rm rb}\right| > Q\left|N_0(f)\right|
\label{eq:transloss}
\end{aligned}
\end{equation}
where $Q>1$ is a required signal-to-noise ratio that ensures the noise interference on ${S}_{\rm 11,b}^{(2)}(f)$ is small. When choosing a low-loss reference plate [$\mathbf  r$], we can assume $\left|T_{\rm r}^{2}\right|\approx 1$. The path loss $\left|A_{\rm b}\right|$ should be small with good system calibration. For practical estimation, let us assume $\left|A_{\rm b}\right| \approx 0.8$. Therefore, we can rewrite the condition as
\begin{equation}
\begin{aligned}
\left|R_{\rm rb}\right| > 1.25Q\left|\frac{N_0}{\left(1-R_{\rm 0r}^{2}\right)}\right|.
\label{eq:reflection2noise}
\end{aligned}
\end{equation}
It can be used to estimate whether a material is measurable. Additionally, when the relative permittivity $\varepsilon_{\rm b}$ of the MUT is large, the reference plate's relative permittivity $\varepsilon_{\rm r}$ should be smaller than $\varepsilon_{\rm b}$. It is worth noting that $\left|{S}_{\rm 11,a}^{(1)}(f)\right| \approx \left|{S}_{\rm 11,b}^{(1)}(f)\right| = \left|A_{\rm b}R_{\rm 0r}\right| > Q\left|N_0(f)\right|$ as well. Consequently, we have the condition:
\begin{equation}
\begin{aligned}
\left|R_{\rm 0r}\right| > 1.25Q\left|N_0\right|.
\label{eq:reflection1noise}
\end{aligned}
\end{equation}

Therefore, when choosing the reference plate and considering the MUT, it is important to fulfill~\eqref{eq:reflection2noise} and~\eqref{eq:reflection1noise}.

\subsection{Permittivity Estimation Workflow}
The following steps outline the workflow for estimating the permittivity using the proposed measurement method:
\begin{enumerate}
\item Measure the environmental reflection coefficient $\hat{S}_{11,0}$.
\item Place the reference plate in the plane wave region and record $\hat{S}_{\rm 11,a}$, where material [$\mathbf  a$] is air.
\item Place the MUT behind the reference plate and record $\hat{S}_{\rm 11,b}$.
\item Remove the MUT and record $\hat{S}^{\prime}_{\rm 11,a}$.
\item Apply~\eqref{eq:newSpa} to calculate $\bar{S}_{11,u}$, where $u$ represents either material [$\mathbf  a$] or [$\mathbf  b$].
\item Use IDFT, e.g., 'ifft' function in \textit{Matlab}, to convert  $\bar{S}_{\rm 11,a}$, $\bar{S}^{\prime}_{\rm 11,a}$ and $\bar{S}_{\rm 11,b}$ into the time domain.
\item Apply the Blackman window or the Kaiser window to extract $\bar{S}_{\rm 11,a}^{(1,1)}(f)$,$\bar{S}_{\rm 11,a}^{(1,2)}(f)$, $\bar{S}_{\rm 11,a}^{(2)}(f)$, $\bar{S}_{\rm 11,b}^{(1)}(f)$, $\bar{S}_{\rm 11,b}^{(2)}(f)$.
\item Apply~\eqref{eq:extract2_2},~\eqref{eq:extract2_3} and~\eqref{eq:extract3} to estimate the phase deviations and inclination angle.
\item Apply~\eqref{eq:meas1} or~\eqref{eq:meas2} to estimate the complex relative permittivity $\varepsilon_{\rm b}$.
\end{enumerate}

\subsection{Simulation Validation}
Full-wave simulations are conducted to validate the proposed method. In the simulation setup, perfect plane waves are assumed, with a free space distance of $d_0 = 200$ mm. The reference plate thickness is set to $W = 30$ mm to ensure clear resolution of the reflected waves from the different interfaces of the reference plate, with a length of $L=10$ mm. The complex permittivity of the reference plate with low loss is chosen as $\varepsilon_{\rm r} = 2-0.02 \rm i$. The inclination angle $\theta_{\rm 0}$ and displacement $dP$ of the reference plate are specified. Considering the high loss property of MUT, $\varepsilon_{\rm b}$ is set to $4-2\rm i$. The frequency range implemented ranged from 130 GHz to 220 GHz. Due to time-gating influences on the upper and down bands, the estimates ranging from 140~GHZ to 210~GHz are considered. The relative error ${err}$ is calculated as $\frac{\left|x-\bar{x}\right|}{x}$, where $\bar x$ represents the estimated values of the real and imaginary parts of the permittivity, and $x$ denotes their true values.
\subsubsection{$\theta_{\rm 0}=0^\circ$ and $dP=0~\rm{\mu m}$}
\begin{figure}[htbp]
    \centering
	\includegraphics[width=0.7\linewidth]{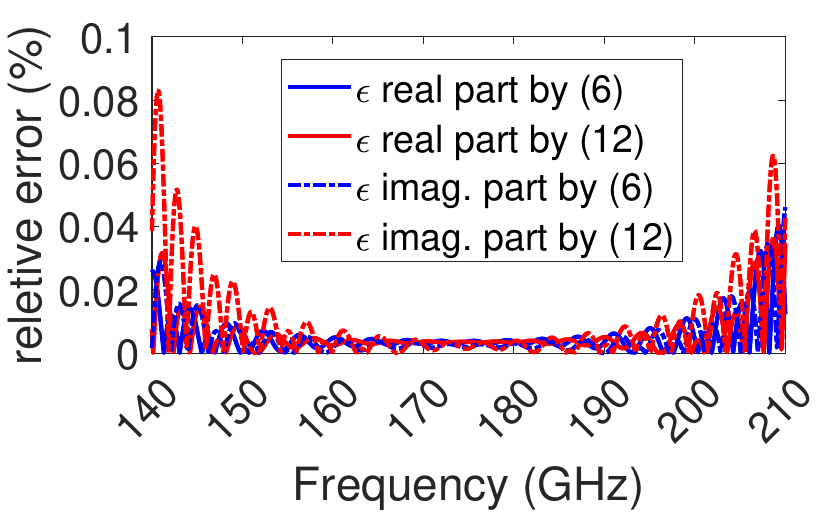}
	\caption{The relative errors of permittivities in the case $\theta_{\rm 0}=0^\circ$ and $dP=0~\rm{\mu m}$.}
	\label{fig:dp=0}
\end{figure}

In this case, we do not include~\eqref{eq:meas2} since its estimates are the same as~\eqref{eq:meas1}. We can see from Fig.~\ref{fig:dp=0} that both formulas can estimate the MUT permittivity accurately. 
\subsubsection{$\theta_{\rm 0}=0^\circ$ and Change $dP$}
\begin{figure}[htbp]
    \centering
	\includegraphics[width=0.7\linewidth]{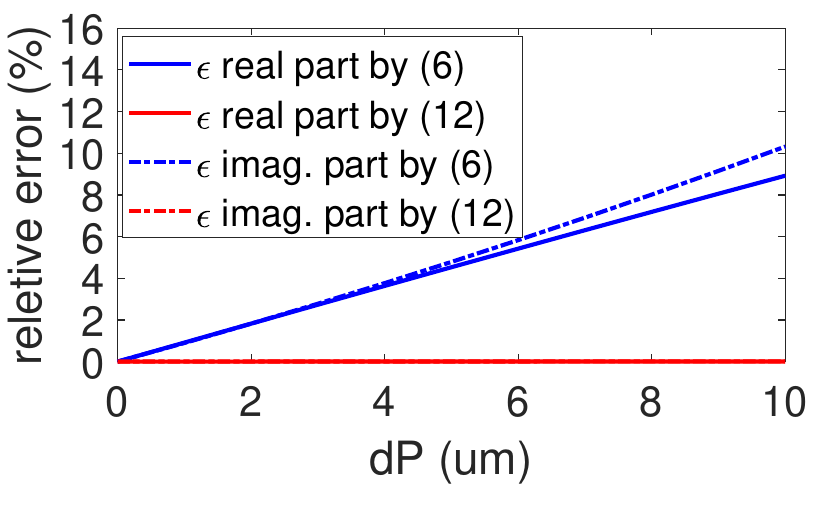}
	\caption{The average relative errors of permittivities in the case $\theta_{\rm 0}=0^\circ$ and $dP\in[0, 10]~\rm{\mu m}$.}
	\label{fig:dpchange}
\end{figure}

In this case, we also do not include~\eqref{eq:meas2}. We change displacement of the reference plate $dP$ from $0~\rm{\mu m}$ to $10~\rm{\mu m}$. The relative errors in Fig.~\ref{fig:dpchange} are calculated by averaging them over the whole frequency range. It can be seen that~\eqref{eq:meas1} keeps the unchanged relative errors while~\eqref{eq:meas} loses accuracy a lot when $dp = 10~\rm{\mu m}$. 
\subsubsection{$\theta_{\rm 0}=1^\circ$ and $dP = 10~\rm{\mu m}$}
\begin{figure}[htbp]
    \centering
	\includegraphics[width=0.7\linewidth]{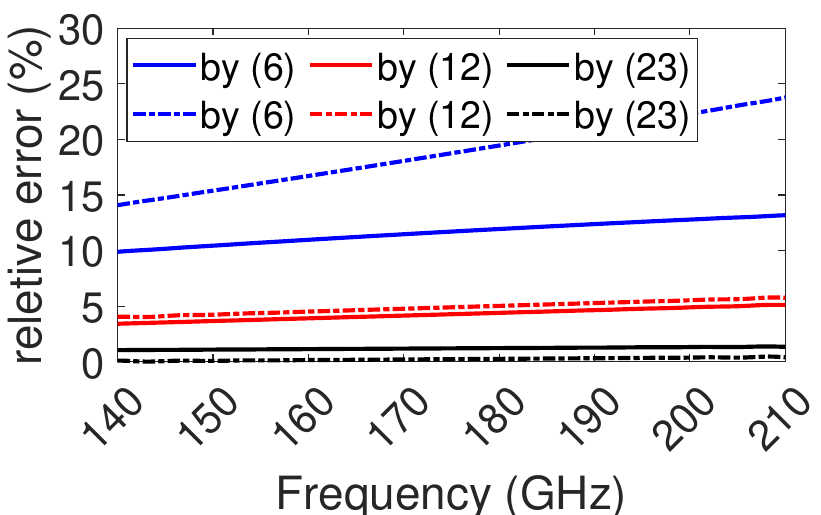}
	\caption{The relative errors of permittivities in the case $\theta_{\rm 0}=1^\circ$ and $dP = 10~\rm{\mu m}$. The solid lines and the dash lines are for real and imaginary parts of permittivities, respectively.}
	\label{fig:theta_1deg}
\end{figure}

In Fig.~\ref{fig:theta_1deg}, it can be seen that~\eqref{eq:meas} loses over $10\%$ accuracy.~\eqref{eq:meas1} can improve the estimate accuracy but it still loses around $4\%$ accuracy.~\eqref{eq:meas1} can further improve the accuracy. 
\subsubsection{$dP = 10~\rm{\mu m}$ and change $\theta_{\rm 0}$}
\begin{figure}[htbp]
    \centering
	\includegraphics[width=0.7\linewidth]{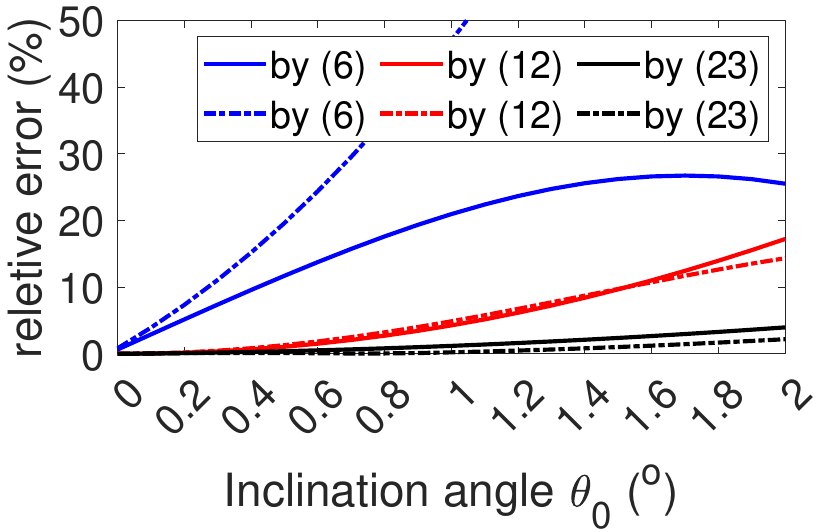}
	\caption{The average relative errors of permittivities in the case $\theta_{\rm 0}\in[0^\circ,2^\circ]$ and $dP = 10~\rm{\mu m}$. The solid lines and the dash lines are for real and imaginary parts of permittivities, respectively.}
	\label{fig:anglechange}
\end{figure}

In this case, we change inclination angle $\theta_{\rm 0}$ from $0~\rm{\mu m}$ to $10~\rm{\mu m}$. The relative errors in Fig.~\ref{fig:anglechange} are calculated by averaging them over the whole frequency range. It can be seen that~\eqref{eq:meas} loses accuracy a lot when $\theta_{\rm 0}>1^\circ$.~\eqref{eq:meas1} can correct phase due to displacement $dP$, but suffers from inclination angle $\theta_{\rm 0}$ significantly when $\theta_{\rm 0}>1^\circ$.~\eqref{eq:meas2} can improve the accuracy because of the inclination angle $\theta_{\rm 0}$. In practice, the estimate errors suffering from the inclination angle $\theta_{\rm 0}$ are far smaller than displacement $dP$. However,~\eqref{eq:meas2} is still useful to improve the accuracy.
\section{Measurement System Setup}
\label{sec:SystemSetup}
\subsection{Introduction to Quasi-optical System}
\label{sec:Quasioptical}
Quasi-optical systems are commonly utilized in frequencies above $100$~GHz for various applications such as imaging and power transmission. These systems have also been used for permittivity measurements~\cite{Kazemipour2015Design, Yashchyshyn2018A, Zhu2021complex}. The advantage of using quasi-optical systems is the generation of a Gaussian beam, which allows for a path loss factor $|A_{\rm b}|\approx 1$, thereby meeting the condition in~\eqref{eq:transloss} more easily.

The quasi-optical system can be optimized in terms of the horn-antenna beamwidth and the number and size of mirrors, depending on the physical size of MUT. According to~\cite{Kazemipour2015Design, Zhu2021complex}, the size of the mirror should be at least three times larger than the beam width when the horn antenna is placed at the focal point distance ($F$) of the parabolic mirror. Hence, the spillover loss and diffraction effects in the plane wave are small. Consequently, the horn antenna should have a narrow beam width to reduce the mirror size and the low side lobes. When measuring the permittivity of a relatively large area, i.e. palm and arm, the plane-wave zone provided by one parabolic mirror can be a proper candidate based on the knowledge in~\cite{Kazemipour2015Design, Zhu2021complex}. However, if we want to measure the permittivity of a relatively small area, i.e. finger, the small quasi-plane-wave zone can be obtained by using a single elliptic mirror or two parabolic mirrors, which can focus the beam and transfer the beam width of the horn antenna to the small beam waist $w_0$. 

For our setup, we selected a corrugated circular conical horn antenna with a gain of more than $20$~ dBi to realize high Gaussicity. We used two identical parabolic off-axis mirrors with a focal point distance of $F = 152.4$~mm and a beam reflection angle of $90^\circ$. With identical focal lengths of the mirrors, the beam waist $W_0$ of the horn is transferred to the target plane.

\begin{figure}[htbp]
    \centering
	\includegraphics[width=0.8\linewidth]{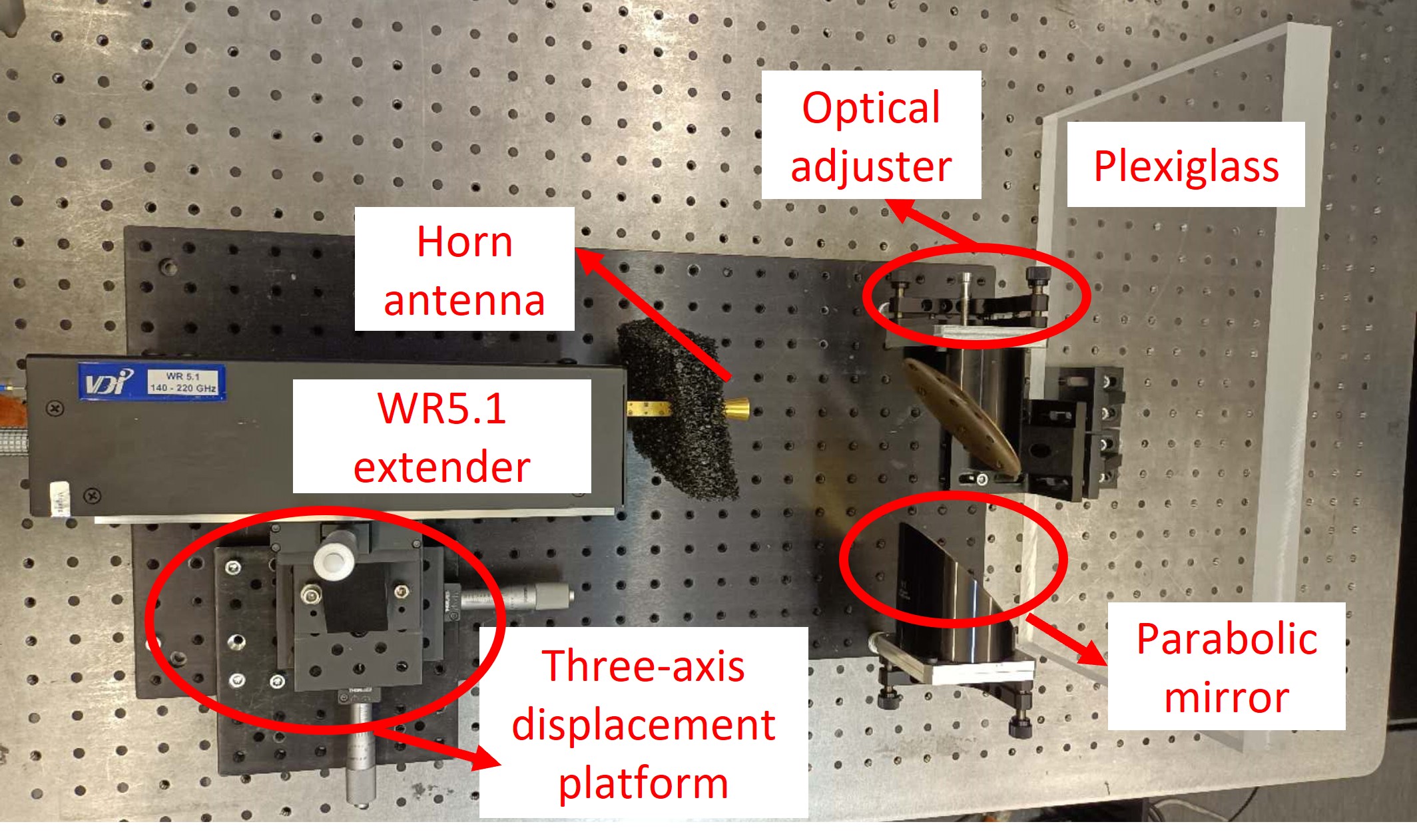}
	\caption{A two-parabolic-mirror quasi-optical system.}
	\label{fig:opticalsystem}
\end{figure}

In the setup, the WR5.1 extender was fixed using combined three-axis optical displacement platforms to find the exact focal point of the mirrors. The optical adjuster was utilized to fix the angles of the mirrors horizontally and vertically. To accurately determine the plane-wave zone of the mirrors, a near-field scanner was employed to check the planes in the location of the reference plate.

\subsection{Quasi-optical System Calibration}
\label{sec:Quasiopticalcali}
Fig.~\ref{fig:opticalsystem} illustrates a two-mirror quasi-optical system. The calibration steps for a two-mirror system can be summarized as follows:

\begin{enumerate}
\item Place the first parabolic mirror and the WR5.1 extender at positions according to the focal-point distance $F$ of the mirror.
\item Scan a probe over a plane of the plane wave zone of the first mirror within the mirror's plane wave zone and adjust the antenna's location based on scanned electric field distributions using three-axis displacement platforms. Also, adjust the mirror's direction using the optical adjuster. Repeat the procedure until a close phase is obtained in the entire plane for different frequencies. Check different planes by moving the near-field scanner along the wave propagation direction.
\item Place the second mirror within the plane wave zone and position the probe of the near-field scanner at the focal point of the second mirror.
\item Scan a probe over a plane at the estimated location of the second mirror's focal point and adjust the direction of the second mirror based on scanned electric field distributions using the optical adjuster.
\item Repeat step 4 until a close phase is achieved within the antenna's beam waist size for different frequencies.
\item Check different planes by moving the near-field scanner along the wave propagation direction to determine the length of the quasi-plane wave zone $z_0$, centered at the focal point.
\end{enumerate}

In general, the above steps ensure quasi-plane wave propagation to illuminate the MUT. For permittivity measurement applications, phase ripples within $<\pm 5^\circ$ in a plane for a wave zone of a plane with one mirror are considered appropriate, which are smaller than $\pm 10^\circ$ realized in~\cite{Kazemipour2015Design}. At the focal point of the second mirror, the phase ripples can be smaller than $<\pm 0.2^\circ$ on the waist plane of the beam. When a metal plate is placed in the plane wave zone, our two-mirror system gave $\hat{S}_{11}>-3.0$~dB at $140$~GHz and $\hat{S}_{11}>-2.5$~dB at $220$~GHz, which is similar to the performance of a four-mirror system with two WR5.1 extenders reported in~\cite{Zhu2021complex}.

The beam waist of our entire system was approximately $w_0 \approx 5$~mm. Based on Gaussian distributions, the radius ($>1.6 w_0$) of the sample under test is capable of catching up to $99.4\%$ power, which sets size requirements for the sample size. According to the relation $z_{\rm c} = \frac{\pi w_0^2}{\lambda}$~\cite{Kazemipour2015Design}, where $z_{\rm c}$ is Rayleigh range, we can estimate the length of the quasi-plane wave zone $ z_0<2z_{\rm c} \approx 74~{\rm mm}$. In our system, $z_0 \approx 12$~ mm based on practical measurement, which is similar to the length of $10$~mm reported in~\cite{Zhu2021complex}, if we consider phase ripples within $<\pm 0.2^\circ$ as a criterion. In addition, we can calculate the confocal distance $2z_{\rm t} = 2\pi\lambda < 13.4~{\rm mm}$, which is the zone that can ensure the beam radius changes little over wave traveling direction.

\subsection{Reference Plate Selection}
\label{sec:PlateSelection}
When measuring the permittivity of human skin, it is necessary for the individual under test to touch their skin on the reference plate in order to flatten the skin surface. It is important that the plate is hard enough so that its deformation is minimal when the skin is touched firmly. We opted for a transparent plexiglass plate as the reference plate since Plexiglass has a reasonable loss factor ($\approx 0.02$), low dielectric constant ($\approx 2.5$)~\cite{afsar1985millimeter}, high hardness, and nearly perfect surface smoothness on the scale of sub-THz wavelength. Due to the high permittivity of human skin, plexiglass can easily satisfy both conditions in~\eqref{eq:reflection2noise} and~\eqref{eq:reflection1noise} simultaneously. In addition, the transparency nature makes it easy to observe the flatness of the skin surface and places clearly during measurements.

\begin{figure}[htbp]
    \centering
	\subfigure[]{
	\includegraphics[width=0.64\linewidth]{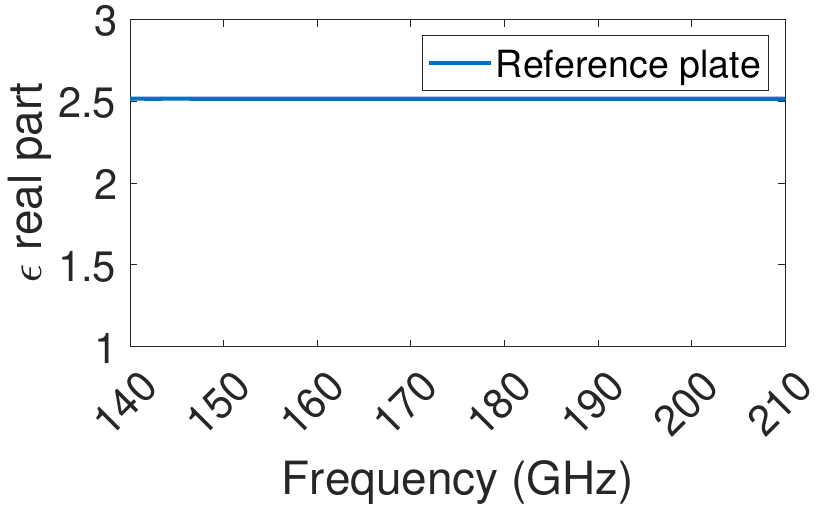}}
 	\subfigure[]{
	\includegraphics[width=0.64\linewidth]{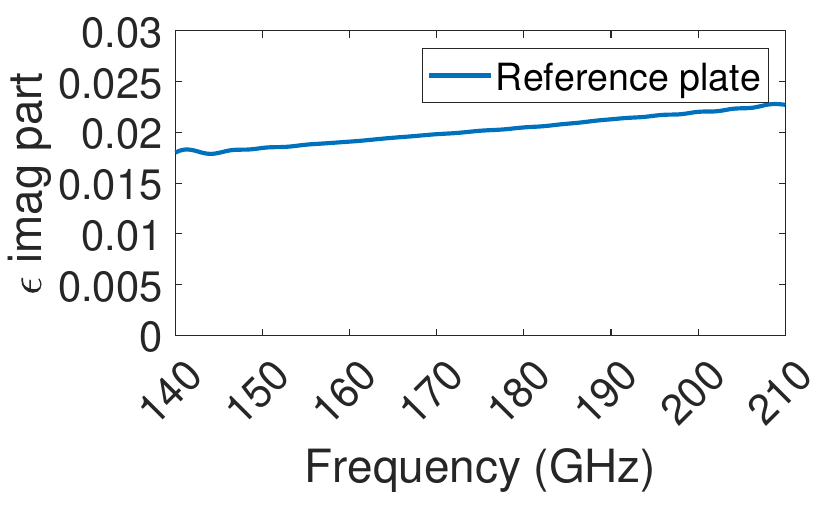}}
	\caption{Measured relative permittivity of the plexiglass: (a) dielectric constant (real part) and (b) loss factor (imaginary part).}
	\label{fig:glassDK}
\end{figure}

Using~\eqref{eq:thickE} and assuming a frequency range of $B_0 = 90$~GHz, we can estimate that the thickness of the plexiglass plate should be $W > 25$~mm. Considering the height of the antenna, a piece of plexiglass with dimensions of $200\times 400\times 30~\rm mm^3$ was chosen. It is important to note that the measured quasi-plane wave zone $z_0$ estimated in Section~\ref{sec:Quasiopticalcali} is only around $12$~mm, which is much shorter than the thickness $W = 30$~mm. It means when we assume that the middle of the reference plate coincides with the focal point of the second mirror, the transmitted wave to the reference plate is not a perfect plane wave in this case. Consequently, we can rewrite $S^{(2)}_{\rm 11,b}$ as
\begin{equation}
\begin{aligned}
S^{(2)}_{\rm 11,b} &= {A_{\rm b}}\left(1-R_{\rm 0r,b}\right)T_{\rm r,b}^{(1)}R_{\rm rb} T_{\rm r,b}^{(2)}\left(1-R_{\rm r0,b}\right){A_{\rm b}^{\prime\prime}} \\& \triangleq U^{(2)} R_{\rm rb},
\label{eq:S211b2new}
\end{aligned}
\end{equation}
where $T_{\rm r,b}^{(1)}$ ($T_{\rm r,b}^{(2)}$) denotes transmission coefficients from Plane C (D) to Plane D (C) shown in Fig.~\ref{fig:permmeasurement}. Similarly, for the known-permittivity material, $S^{(2)}_{\rm 11,a}$ can be expressed as
\begin{equation}
\begin{aligned}
S^{(2)}_{\rm 11,a} = U^{(2)} R_{\rm ra}.
\label{eq:S211a2new}
\end{aligned}
\end{equation}
Therefore, we can find that our calculation for $R_{\rm rb}$, given by $R_{\rm rb} = R_{\rm ra}\frac{S^{(2)}_{\rm 11,b}}{S^{(2)}_{\rm 11,a}}$, only depends on interface reflections. The open angle of the beam can be estimated as $\theta_0 = \frac{\lambda}{\pi w_0} < 8^\circ$. We can approximate that the effective refraction angle $\bar{\theta}_{\rm t} < 3^\circ$ according to Snell’s law. Hence, we have
\begin{equation} 
R_{\rm rb} = \frac{\cos{\theta_{\rm t}} - \sqrt{\frac{\varepsilon_{\rm b}}{\varepsilon_{\rm r}} - \sin{\theta_{\rm t}^2}} }{\cos{\theta_{\rm t}} + \sqrt{\frac{\varepsilon_{\rm b}}{\varepsilon_{\rm r}} - \sin{\theta_{\rm t}^2}}} \approx \frac{\sqrt{\varepsilon_{\rm r}} -\sqrt{\varepsilon_{\rm b}}}{\sqrt{\varepsilon_{\rm r}}+\sqrt{\varepsilon_{\rm b}}}. 
\label{eq:rrr}
\end{equation}
The difference between the approximation and the exact value can be smaller than the measurement uncertainty when properly set up. Thus, the influence of non-plane wave propagation on the measurement results can be neglected. In Section~\ref{sec:thickness}, we show that the above assumption is rational using measurements. In addition, the permittivity of the reference plate (plexiglass) was measured using the method in~\cite{xue2024thick}, shown in Fig.~\ref{fig:glassDK}.
The frequency range shown is $140-210$~GHz due to the truncation distortion caused by time gating in the lowest and highest frequency parts of the S-parameters. The measured permittivity is expected to have low and stable values for both the real and imaginary parts, making it a suitable reference plate for human skin permittivity measurements.

\section{Experimental Measurement and Analysis}
\label{sec:MeasurementResults}
We calibrated the system according to the method introduced in Section~\ref{sec:Quasiopticalcali}. The initial state, including inclination and location, of plexiglass was calibrated by ensuring a maximum reflection coefficient ($S^{(1)}_{\rm 11,a}(f)$) at the WR5.1 extender side. In this setup, the frequency range was set from $130$~GHz to $220$~GHz, and the intermediate frequency was $\rm IF = 1 \rm ~kHz$. We used $1601$ sampling points to reduce IDFT aliasing, and the length of the Blackman window in time gating was $60\delta_{\rm t}$.

\subsection{Permittivity Measurement of Copy Papers}
To verify the reliability of the measurement results obtained using this approach, we conducted measurements on stacked A4 copy papers ($5~\rm cm$ thickness). The copy papers are flat and soft, allowing them to easily make contact with the plexiglass without any gaps. The relative permittivity of the reference complex of the papers is reported as $\varepsilon_{\rm r} \approx 2.8-0.15\rm i$ at $50\rm~GHz$ and $\varepsilon_{\rm r} \approx 2.5-0.15\rm i$ at $75$~GHz in~\cite{wang2022fast}. Therefore, we estimated the complex permittivity to be approximately $\varepsilon_{\rm r} \approx 2.1-0.15\rm i$ at sub-THz frequencies based on linear extrapolation along frequencies.
\begin{figure}[htbp]
    \centering
	\subfigure[]{
	\includegraphics[width=0.64\linewidth]{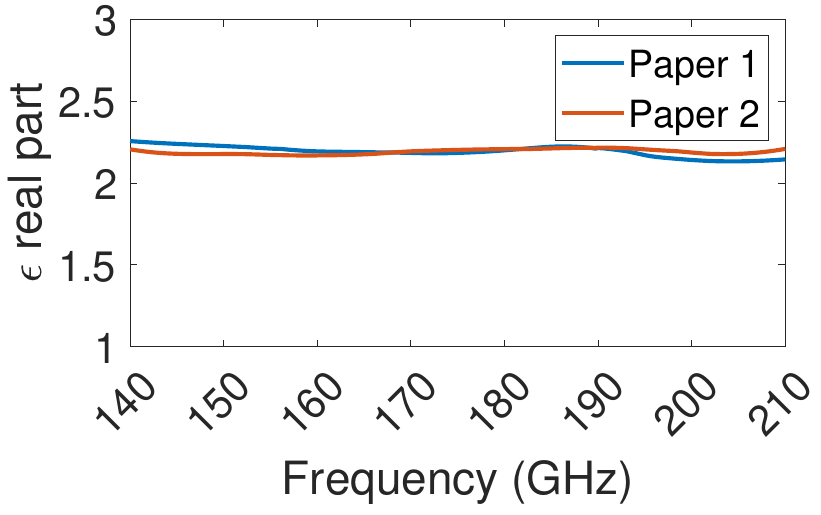}}
 	\subfigure[]{
	\includegraphics[width=0.64\linewidth]{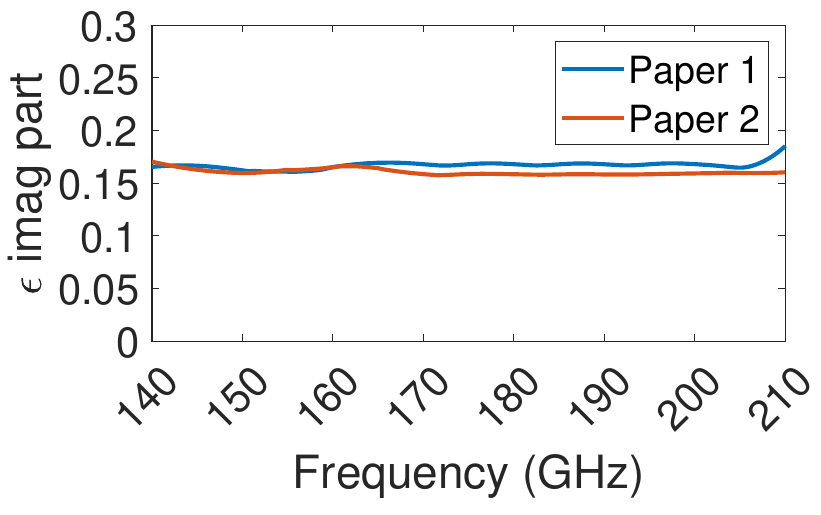}}
	\caption{Measured complex permittivity of two paper samples: (a) dielectric constant (real part) and (b) loss factor (imaginary part).}
	\label{fig:paperDK}
\end{figure}
Since the displacement and inclination effects during the paper measurements are negligible, the estimates obtained by~\eqref{eq:meas},~\eqref{eq:meas1} and~\eqref{eq:meas2} are nearly the same, so that we only show the measured results obtained by~\eqref{eq:meas2} in Fig.~\ref{fig:paperDK}. It can be seen that the two paper samples exhibit similar permittivity characteristics. The real part shows a difference of $<0.1$ between the two samples, while the imaginary part exhibits a difference of $0.02$ for most frequencies. These values are consistent with the estimate $\varepsilon_{\rm r} \approx 2.1-0.15\rm i$. 
\subsection{Human Skin Permittivity Measurement Analysis}
\begin{figure}[htbp]
    \centering
	\includegraphics[width=0.75\linewidth]{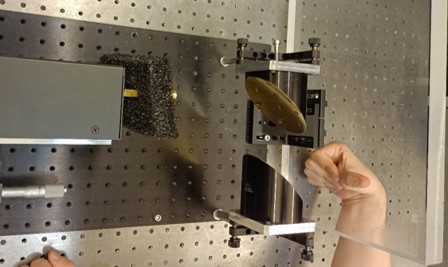}
	\caption{The two-mirror measurement setups for human skins}
	\label{fig:setups}
\end{figure}

Human fingers', palms', and arms' skin permittivities were measured. The setups are introduced first as shown in Fig.~\ref{fig:setups}. 
\subsubsection{Phase Correction Analysis}
\begin{figure}[htbp]
    \centering
	\subfigure[]{
	\includegraphics[width=0.64\linewidth]{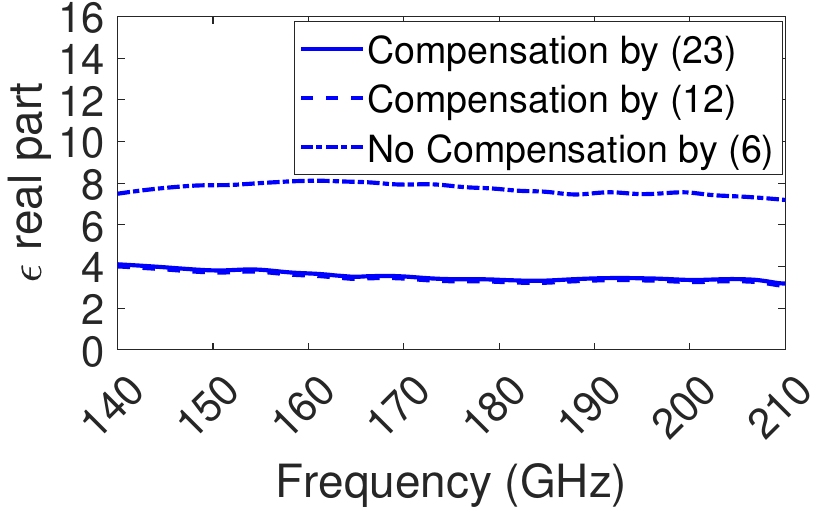}}
 	\subfigure[]{
	\includegraphics[width=0.64\linewidth]{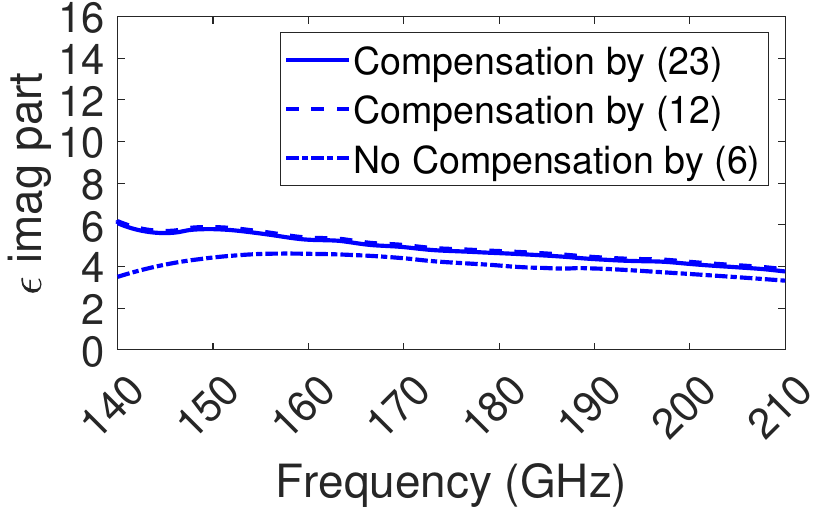}}
	\caption{Measured relative permittivity of the human arm skin with compensation and without phase correction: (a) dielectric constant (real part) and (b) loss factor (imaginary part).}
	\label{fig:compensation}
\end{figure}

This section demonstrates how the phase correction formulas,~\eqref{eq:meas1} for displacement and~\eqref{eq:meas2} for displacement and inclination of the reference plate, can improve the measurement results. Fig.~\ref{fig:compensation} illustrates the estimates with human arm. When displacement is not negligible,~\eqref{eq:meas} does not provide a meaningful value for permittivity, resulting in a large discrepancy, according to the knowledge shown in~\cite{pickwell2004vivo, zakharov2009full} that the relative permittivity of the human skin should be close to that of the epidermis layer ($3.3-5.2 \rm i$) since the stratum corneum layer is only $\sim 15~\rm \mu m$ on the arm. Comparing the curves obtained using~\eqref{eq:meas1} and~\eqref{eq:meas2}, it can be observed that~\eqref{eq:meas2} only compensates for approximately $0.1$ in both real and imaginary parts. This is because the reference plate used in the measurement is thick and rigid, resulting in minimal angle mismatch. However,~\eqref{eq:meas2} is still useful for improving the accuracy of higher frequency measurements. In the following sections, permittivity estimates are based on~\eqref{eq:meas2}.
\subsubsection{Reference Plate's Thickness Influences}
\label{sec:thickness}
\begin{figure}[htbp]
    \centering
	\subfigure[]{
	\includegraphics[width=0.64\linewidth]{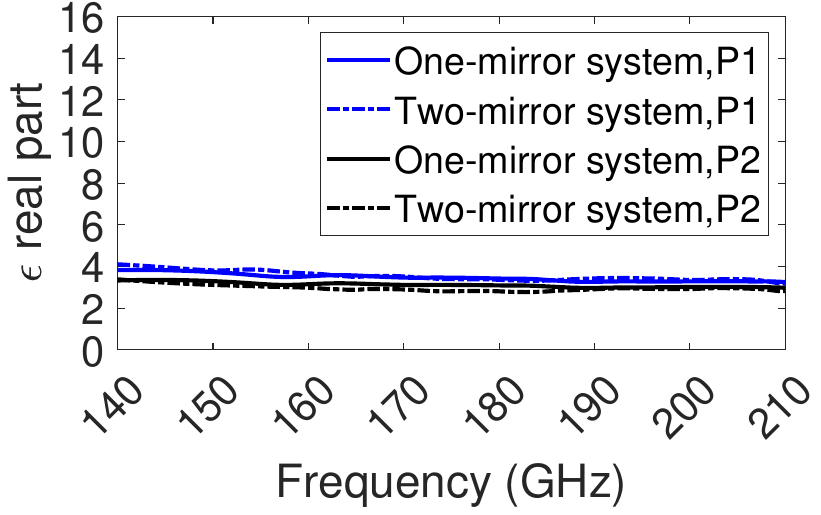}}
 	\subfigure[]{
	\includegraphics[width=0.64\linewidth]{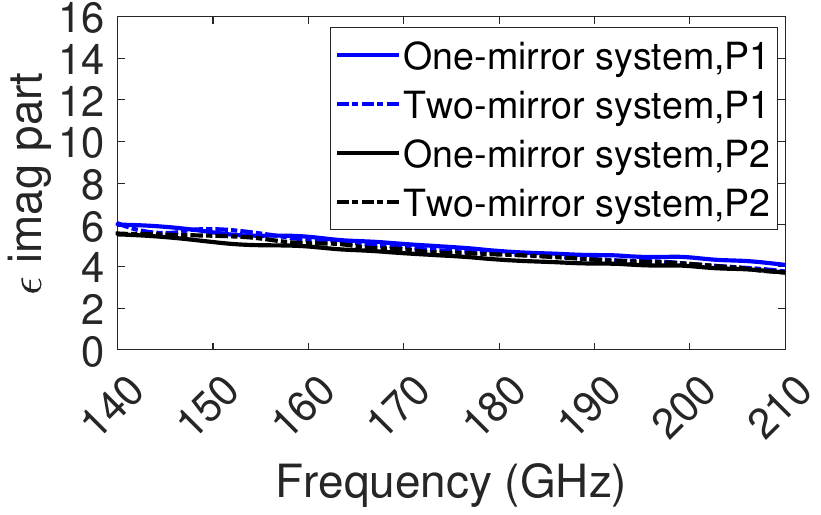}}
	\caption{Measured relative permittivity of human arm skin by one-mirror and two-mirror systems with a $30-\rm mm$ thick reference plate: (a) dielectric constant (real part) and (b) loss factor (imaginary part).}
	\label{fig:onetwo}
\end{figure}

In Section~\ref{sec:PlateSelection}, we have proved that $30-\rm mm$ thick reference plate in the two-mirror system can still provide an accurate enough permittivity characterization by theoretical estimation. Besides, the evidence given by measurement results presented in Fig.~12 of~\cite{Zhu2021complex} shows this measurement system can estimate permittivity accurately for MUT thicker than $20$~mm. Here, in order to further illustrate that $30-\rm mm$ thickness of the reference plate only affects permittivity characterizations a little, we conducted measurements using the one-mirror and two-mirror systems with two individuals on the same day, ensuring consistency in their arm moisture statuses. The plane-wave calibration for the one-mirror system was implemented by following the calibration instruction (steps: 1-2) in Section~\ref{sec:Quasiopticalcali}. As noted above, the one-mirror system provides a large plane-wave zone, i.e., $z_0 = 8$~cm~\cite{Kazemipour2015Design, Zhu2021complex}. Fig.~\ref{fig:onetwo} shows the comparison results between the one-mirror and two-mirror systems. It can be observed that the average differences between the two systems are smaller than $0.3$ for both the real and imaginary parts of the permittivity. These differences can mainly be attributed to the fact that the one-mirror system has a wider plane wave zone. Moreover, the measured regions are not exactly the same. Therefore, we can further conclude that the $30\rm mm$ thickness reference plate can provide adequate accurate measurement results.
\subsubsection{Sweat Influences on Measurement Results}
Sweat makes the human skin present a different permittivity~\cite{Zhekov2019}. To account for the effects of sweat on the measurement results, we chose finger permittivity as an example, since sweat is easily on the finger and palm but rarely on the arm. It is noted that only thumbs can be accurately measured since other fingers do not cover the beam waist $w_0 = 5$~mm of the two-mirror system.
\begin{figure}[htbp]
    \centering
	\subfigure[]{
	\includegraphics[width=0.64\linewidth]{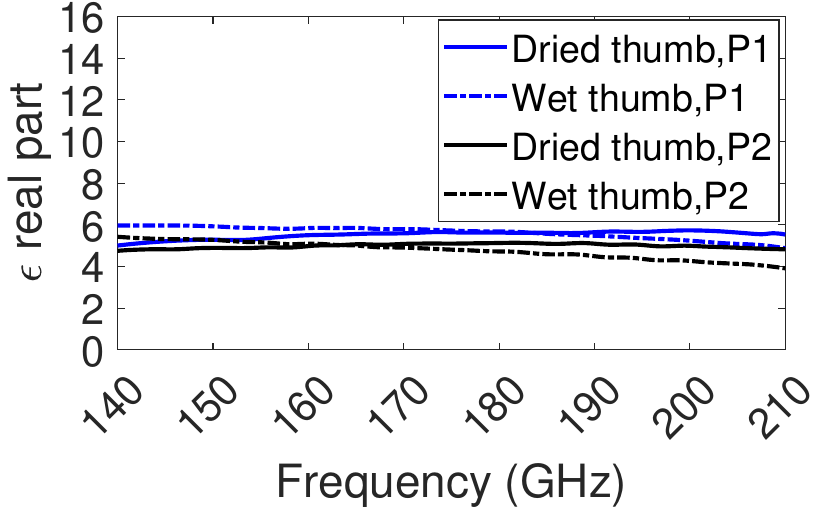}}
 	\subfigure[]{
	\includegraphics[width=0.64\linewidth]{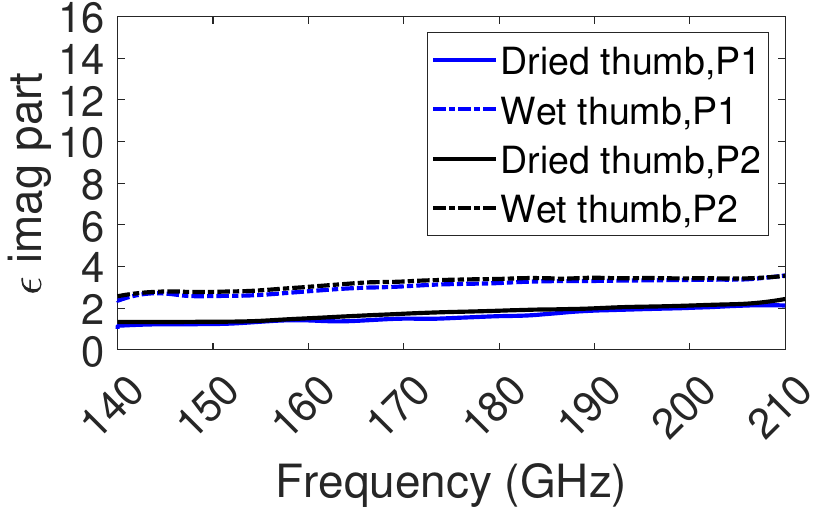}}
	\caption{Measured permittivity of one-mirror and two-mirror system with a $30-\rm mm$ thick reference plate: (a) dielectric constant (real part) and (b) loss factor (imaginary part).}
	\label{fig:sweat}
\end{figure}
We conducted measurements on dry thumbs and then repeated the measurements after moisturizing the thumbs to simulate sweat effects. The results are shown in Fig.~\ref{fig:sweat}. It can be observed that the real part of the permittivity for the dry thumb remains relatively constant across the frequency range. However, the wet thumb shows a decreasing trend as the frequency increases. In terms of the imaginary part, the wet thumb demonstrates higher conductivity compared to the dry thumb. The difference between dry and wet thumbs is approximately $1$ for both real and imaginary parts. These findings highlight the impact of sweat on the permittivity estimates of human fingers.

\begin{figure}[htbp]
    \centering
	\subfigure[]{
	\includegraphics[width=0.64\linewidth]{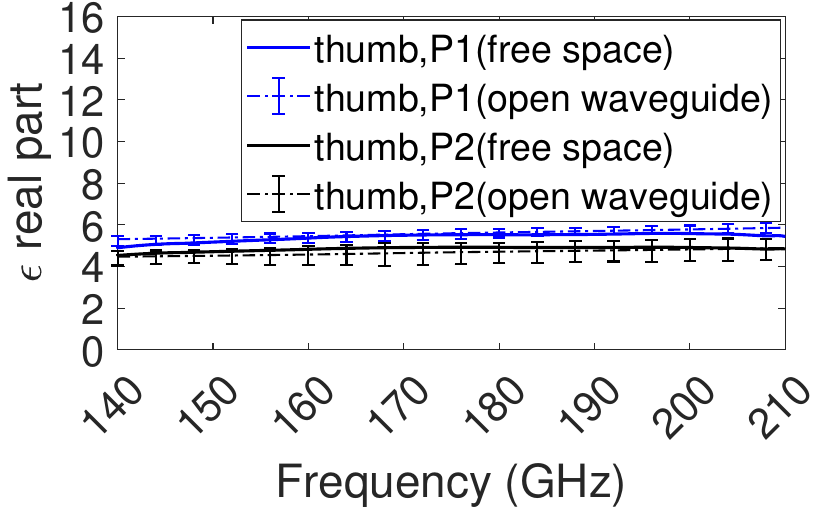}}
 	\subfigure[]{
	\includegraphics[width=0.64\linewidth]{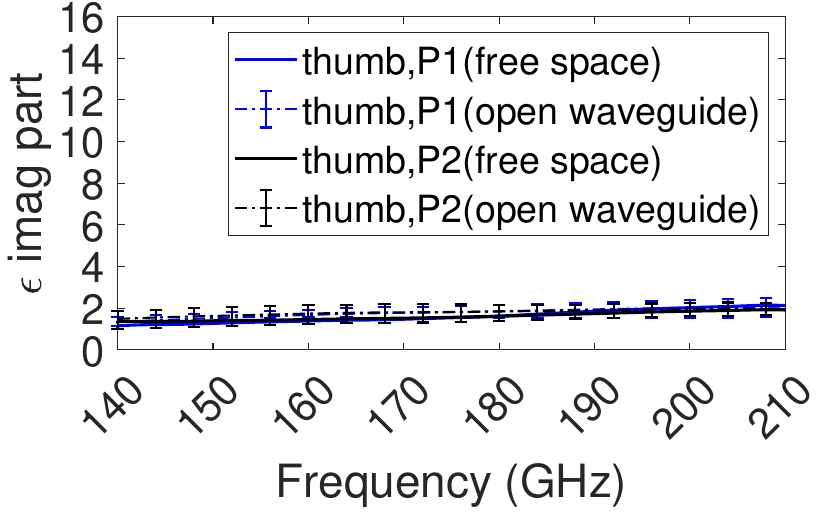}}
	\caption{Measured relative permittivity by using the proposed method (free space) and the method in~\cite{xue2024Human} (open waveguide): (a) dielectric constant (real part) and (b) loss factor (imaginary part). The error bars are defined by the maximum deviations of permittivity estimates compared with mean values.}
	\label{fig:twomethod}
\end{figure}
\subsection{Comparisons with the Method in~\cite{xue2024Human}}
In order to further validate the proposed method, we compare the permittivity estimates with those from a different method, elaborated in~\cite{xue2024Human}. Differing from this present method, the one in~\cite{xue2024Human} can only characterize permittivity within a small area of a MUT, defined by the aperture of the WR5 waveguide of $\sim 0.8~{\rm mm^2}$, so we can name this permittivity as `local permittivity’. In contrast, the present method (two-mirror system) can characterize human skin’s permittivity within a relatively large area of $\sim 110{\rm mm^2}$, defined by the beam waist of the quasi-optical system, so we can name this permittivity as `effective permittivity'. In this experiment, we used these two methods to characterize the thumb permittivity for two persons within two hours, which can ensure that their thumb's permittivity is similar to a large extent when measured by these two methods. In order to estimate the local mean permittivity, we measured the permittivity in the random area of the thumb 20 times using the method in~\cite{xue2024Human} and derived the maximum deviation of the permittivity estimates. The setup for the present method is the same as that shown in Fig.~\ref{fig:setups}. The permittivity characterized by the two methods is shown in Fig.~\ref{fig:twomethod}, showing that they have $<10\%$ difference for the real part of the permittivity, while they have $<20\%$ difference for the imaginary part for the frequency range.

\begin{table*}[htbp]
	\begin{center}
		\caption{Measurement Uncertainty Estimations at $140\rm~GHz$} 
		\label{uncertainty}
  \begin{tabular}{|c|c|c|c|c|c|c|c|c|}\hline &\multirow{2}{*}{$\Delta (S_{11}^{(2)})^{\rm dB}$}&\multirow{2}{*}{$\Delta (\angle S_{11}^{(2)})^{\circ}$}&\multicolumn{4}{c|}{\eqref{eq:meas}}&\multicolumn{2}{c|}{\eqref{eq:meas2}}
  \\ \cline{4-9}
   & & &$\Delta$DK&$\Delta$LF&$\Delta$DK$_{\rm dev}$&$\Delta$LF$_{\rm dev}$&$\Delta$DK&$\Delta$LF\\\hline
    paper&$\pm 0.05$&$\pm 0.25$&$\pm 0.24\%$&$\pm 2.06\%$&$0.95\%$&$0.87\%$ &$\pm 0.17\%$&$\pm 1.76\%$\\
    thumb&$\pm 0.12$&$\pm 1.0$&$\pm 0.56\%$&$\pm 3.35\%$&$34.5\%$&$10.5\%$ &$\pm 0.41\%$&$\pm 1.26\%$\\    
    palm&$\pm 0.20$&$\pm 0.75$&$\pm 1.14\%$&$\pm 2.55\%$&$46.5\%$&$10.3\%$ &$\pm 0.57\%$&$\pm 0.98\%$\\    
    arm&$\pm 0.22$&$\pm 2.0$&$\pm 6.67\%$&$\pm 3.34\%$&$92.2\%$&$17.8\%$ &$\pm 1.33\%$&$\pm 1.04\%$\\ \hline   
 	\end{tabular}
	\end{center}
   $\Delta$DK represents the relative uncertainty of the real part of the complex permittivity; $\Delta$DK$_{\rm dev}$ represents the relative deviation of the real part of the complex permittivity between~\eqref{eq:meas} and~\eqref{eq:meas2}. 
   $\Delta$LF represents the relative uncertainty of the imaginary part of the complex permittivity;
   $\Delta$LF$_{\rm dev}$ represents the relative deviation of the imaginary part of the complex permittivity between~\eqref{eq:meas} and~\eqref{eq:meas2}. 
   ~\eqref{eq:meas} represents no phase correction;~\eqref{eq:meas2} represents apply phase correction; the values are obtained by averaging the uncertainties for thumb, palm, and arm permittivity measurements across all individuals.
	\end{table*}
  \begin{figure*}[htbp]
    \centering
	\subfigure[]{
	\includegraphics[width=0.32\linewidth]{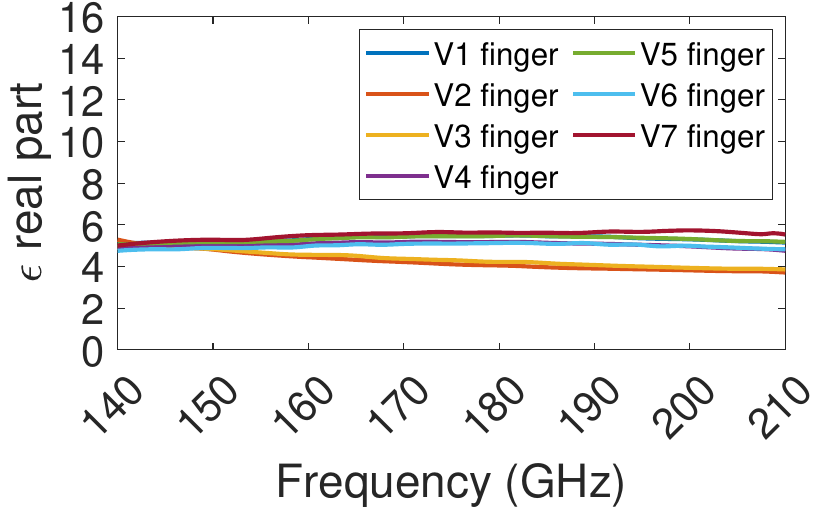}}
  \subfigure[]{
	\includegraphics[width=0.32\linewidth]{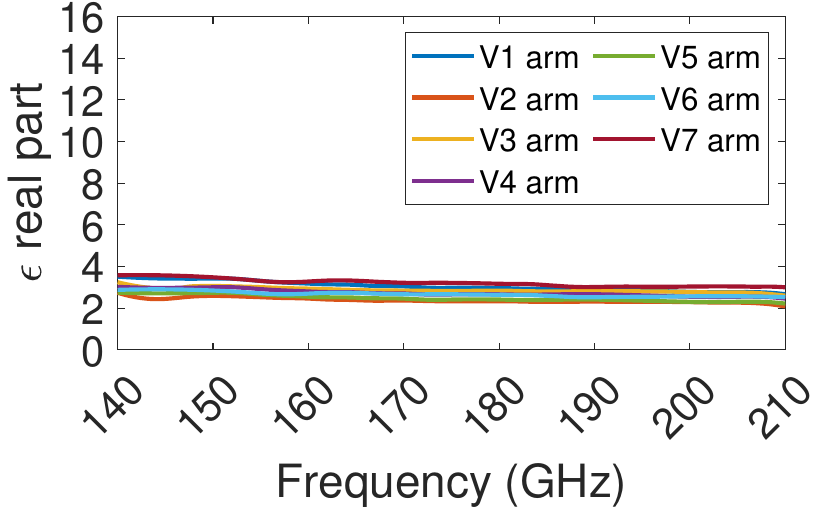}}
 \subfigure[]{
	\includegraphics[width=0.32\linewidth]{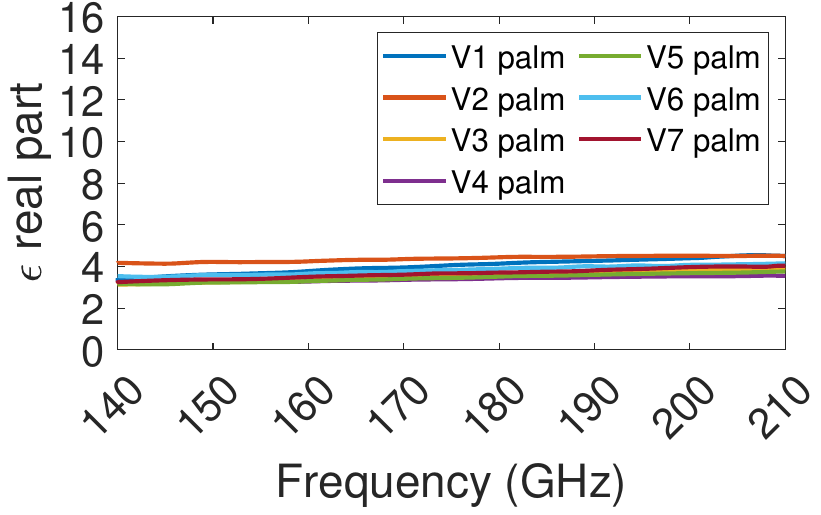}}
 \subfigure[]{
	\includegraphics[width=0.32\linewidth]{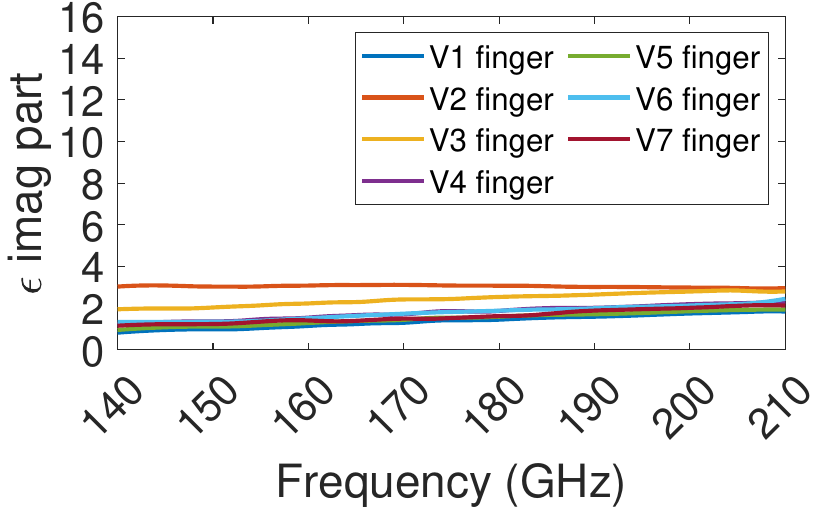}}
  	\subfigure[]{
	\includegraphics[width=0.32\linewidth]{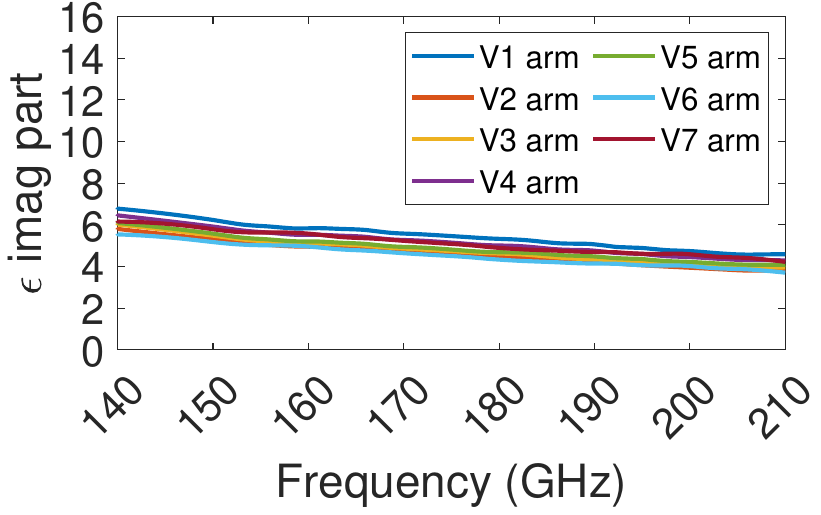}}
 	\subfigure[]{
	\includegraphics[width=0.32\linewidth]{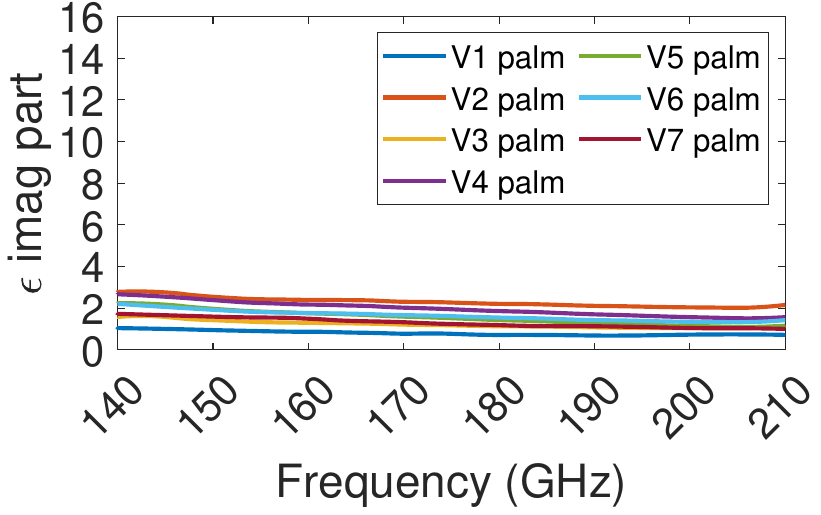}}
	\caption{Measured relative permittivity of different persons' fingers: (a) dielectric constant (real part) and (d) loss factor (imaginary part); the measured permittivity of different persons' arms: (b) dielectric constant (real part) and (e) loss factor (imaginary part); the measured permittivity of different persons' palms: (c) dielectric constant (real part) and (f) loss factor (imaginary part).}
	\label{fig:finarmpalm}
\end{figure*}

\subsection{Measurement Uncertainty Analysis}
When measuring the permittivity of human skin, it is important to consider the uncertainty not only of the measurement system but also of how human tremor affects the measured S parameters. The VNA measurement uncertainty can be estimated using the approach described in~\cite{Zhu2021complex} or calculated using statistical methods. However, accurately quantifying the uncertainty due to human tremor is challenging. Therefore, we calculate the estimate uncertainties by computing the relative uncertainty of several measurements $\hat{\sigma_{\rm s}}$, which is defined by
\begin{equation}
 \hat{\sigma_{\rm s}} = \frac{\sigma_{\rm s}}{\bar x},  
\end{equation}
where $\sigma_{\rm s}$ is the standard deviation of these repeated measurements, and $\bar x$ represents the mean value of the real or imaginary parts of permittivities.
In TABLE~\ref{uncertainty}, we present the estimated VNA uncertainty $\Delta (S_{11}^{(2)})^{\rm dB}$ and $\Delta (\angle S_{11}^{(2)})^{\circ}$ for paper and human skin permittivity measurements at $140\rm~GHz$, which can be a representative for all frequencies. These uncertainties were calculated based on 20 measurements taken within a 20-second interval.

From the data in TABLE~\ref{uncertainty}, we can observe that the effect of arm tremor measured $S_{11}^{(2)}$ is larger than that of thumb and palm. Phase correction, as indicated by~\eqref{eq:meas2}, can significantly reduce the influence of tremors on the overall permittivity uncertainties. When it comes to paper and lens, there is only a small improvement for compensated permittivity uncertainties when comparing the columns representing phase correction to those without phase correction. This proposed method shows a similar uncertainty of $\pm 1\%$ to~\cite{xue2024Human} smaller than that reported in~\cite{gao2018towards} when characterizing the permittivity of human skin.

\subsection{Human Skin Permittivity Results}
Permittivity estimates of human skin were obtained from seven volunteers. They were instructed not to clean or dry their hands and arms to capture the daily status and natural condition of human skin. The permittivity estimates presented in Fig.~\ref{fig:finarmpalm} show distinct characteristics in the permittivity of different regions of the human skin. Apart from the second and the third volunteers, the permittivity estimates of fingers show different trends from the others, indicating the presence of sweat on those individuals' fingers shown in Fig.~\ref{fig:finarmpalm} (a) and (d). The low imaginary-part values of the finger permittivity are attributed to the thick stratum corneum, which obviously has different characteristics compared to other layers of the skin. In contrast, the arm permittivities in Fig.~\ref{fig:finarmpalm} (b) and (e) exhibit a larger imaginary part, suggesting a higher conductivity. This is consistent with the fact that the stratum corneum is thinner on the arm compared to the fingers, and the permittivity is closer to that of water. Interestingly, the permittivity of the arm region is more consistent across individuals. On the other hand, the palm permittivities exhibit larger variations compared with arms, shown in Fig.~\ref{fig:finarmpalm} (c) and (f), which could be attributed to variations of, e.g., hydration levels and skin thickness, among others. The complex permittivities of the finger, palm, and arm of seven volunteers show standard deviations by $<0.95$, $<0.46$ and $<0.53$ from $140$~GHz to $210$~GHz, respectively.
\begin{figure}[htbp] 
    \centering
	\includegraphics[width=0.64\linewidth]{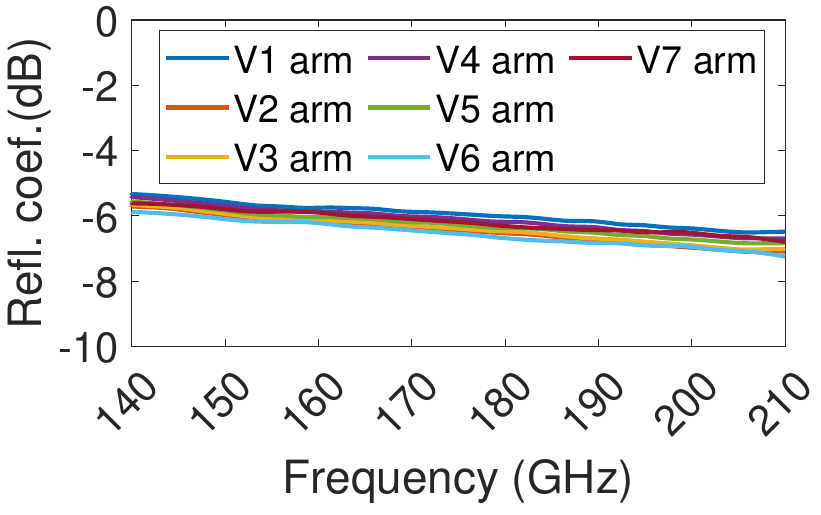}
	\caption{Reflection coefficients in free space using the permittivity shown in Fig.~\ref{fig:finarmpalm}(b) and (e).}
	\label{fig:reflection}
\end{figure}

The reflection coefficients in free space, calculated based on arms' permittivities and shown in Fig.~\ref{fig:reflection}, illustrate that their variations are smaller than $0.6\rm~dB$ at $140\rm~GHz$ and $0.8\rm~dB$ at $210\rm~GHz$. The small variation implies that the human-skin differences between individuals may be neglected in human-electromagnetic-wave interaction studies in sub-THz radio communications. 

\section{Conclusions}
\label{sec:conclusion}
This paper has proposed a measurement system for accurately characterizing the effective permittivity of human skin for different regions of the human body at sub-THz. The effectiveness of the proposed method has been demonstrated through the determination of the permittivity of human fingers, palms, and arms. Uncertainty of the obtained permittivity estimates with the phase correction methods was within $\pm 1.33~\%$, which is mainly due to the unavoidable effects of human tremors. Comparing the proposed method with the existing one in~\cite{xue2024Human}, the permittivity estimates showed $<10~\%$ and $<20~\%$ differences for the real and imaginary parts, respectively, over $140$ to $210$~GHz range. Complex permittivities of the arm, palm, and thumb of seven volunteers have shown significant variations. Sweat had a dominant influence on permittivity in the frequency range, leading to an even over $100~\%$ increase for the loss factor according to the measured palm permittivities. Therefore, future simulation studies on human-electromagnetic-wave interactions in the sub-THz bands require the use of human models where varying complex permittivities are assigned to different parts of the human body if high accuracy is needed.

\section*{Acknowledgments}
All authors thank those volunteers engaged in finger-permittivity measurements in the School of Electrical Engineering of Aalto University.

\bibliographystyle{IEEEtran}
\footnotesize
\bibliography{refrealhand}
\end{document}